\documentclass[12pt,preprint]{aastex}
 
 \usepackage{epstopdf}
 \usepackage{longtable}
\usepackage{float}

 \slugcomment{Accepted}

\newcommand\pv{\mbox{$p_{V}$}}
\newcommand\pIR{\mbox{$p_{IR}$}}
\newcommand\irfactor{\mbox{$p_{IR}/p_{V}$}}

\begin{document}

 \DeclareGraphicsExtensions{.pdf,.gif,.jpg,.png}

 \title{Initial Performance of the NEOWISE Reactivation Mission}
\author{A. Mainzer\altaffilmark{1}, J. Bauer\altaffilmark{1,2},  R. M. Cutri\altaffilmark{2}, T. Grav\altaffilmark{3}, J. Masiero\altaffilmark{1}, R. Beck\altaffilmark{2}, P. Clarkson\altaffilmark{4}, T. Conrow\altaffilmark{2}, J. Dailey\altaffilmark{2}, P. Eisenhardt\altaffilmark{1}, B. Fabinsky\altaffilmark{1}, S. Fajardo-Acosta\altaffilmark{2}, J. Fowler\altaffilmark{2}, C. Gelino\altaffilmark{2}, C. Grillmair\altaffilmark{2}, I. Heinrichsen\altaffilmark{1}, M. Kendall\altaffilmark{4}, J. Davy Kirkpatrick\altaffilmark{2}, F. Liu\altaffilmark{1}, F. Masci\altaffilmark{2}, H. McCallon\altaffilmark{2}, C. R. Nugent\altaffilmark{1},  M. Papin\altaffilmark{2}, E. Rice\altaffilmark{1},  D. Royer\altaffilmark{1}, T. Ryan\altaffilmark{4}, P. Sevilla\altaffilmark{5}, S. Sonnett\altaffilmark{1}, R. Stevenson\altaffilmark{1}, D. B. Thompson\altaffilmark{5}, S. Wheelock\altaffilmark{2}, D. Wiemer\altaffilmark{4}, M. Wittman\altaffilmark{2}, E. Wright\altaffilmark{6}, L. Yan\altaffilmark{2}}

 \altaffiltext{1}{Jet Propulsion Laboratory, California Institute of Technology, Pasadena, CA 91109 USA}
 \altaffiltext{2}{Infrared Processing and Analysis Center, California Institute of Technology, Pasadena, CA 91125, USA}
\altaffiltext{3}{Planetary Science Institute, Tucson, AZ USA}
\altaffiltext{4}{Ball Aerospace and Technology Center, Boulder, CO USA}
\altaffiltext{5}{Space Dynamics Laboratory, Utah State University, Logan, UT USA}
\altaffiltext{6}{Department of Physics and Astronomy, UCLA, PO Box 91547, Los Angeles, CA 90095-1547 USA}
 \email{amainzer@jpl.nasa.gov}

 \begin{abstract}
NASA's Wide-field Infrared Survey Explorer (WISE) spacecraft has been brought out of hibernation and has resumed surveying the sky at 3.4 and 4.6 $\mu$m.  The scientific objectives of the NEOWISE reactivation mission are to detect, track, and characterize near-Earth asteroids and comets.  The search for minor planets resumed on December 23, 2013, and the first new near-Earth object (NEO) was discovered six days later.  As an infrared survey, NEOWISE detects asteroids based on their thermal emission and is equally sensitive to high and low albedo objects; consequently, NEOWISE-discovered NEOs tend to be large and dark.  Over the course of its three-year mission, NEOWISE will determine radiometrically-derived diameters and albedos for $\sim$2000 NEOs and tens of thousands of Main Belt asteroids.  The 32 months of hibernation have had no significant effect on the mission's performance.  Image quality, sensitivity, photometric and astrometric accuracy, completeness, and the rate of minor planet detections are all essentially unchanged from the prime mission's post-cryogenic phase.    

 \end{abstract}

 \section{Introduction}
Understanding the numbers, orbits, and physical properties of the asteroids and comets that approach Earth is essential both for characterizing the population of objects that pose a potential impact hazard, as well as for planning an appropriate mitigation strategy should one be discovered on a threatening trajectory.  Of the approximately 10,700 near-Earth objects (NEOs; asteroids and comets with perihelia less than 1.3 AU) discovered to date, only the most basic properties (orbital parameters and absolute magnitude H) are known for all but $\sim$2000 at present.  Well-determined physical measurements such as taxonomic classification, sizes, and shapes and rotational states are being determined for $\sim$100 additional NEOs each year \citep[e.g. the MIT-UH-IRTF Joint Campaign for NEO Spectral Reconnaisance,][and many others]{Xu.1995a, Tedesco.2002a, Benner.2008a, Durech.2010a, Reddy.2010a}.  Much remains to be learned about the detailed physical properties of the NEOs, particularly since the $\sim$10,700 known NEOs represent only a small fraction of the total population at all size ranges.  Recent estimates suggest that there are 20,500$\pm$3000 near-Earth asteroids (NEAs) larger than 100 m in diameter; it is estimated that only $\sim$25\% of these have been discovered to date \citep{Mainzer.2011b}.  For sizes smaller than 100 m, survey completeness drops precipitously.    

Because impact energy is proportional to diameter cubed for a given density, relatively small errors in diameter can lead to large errors in predicted impact energy.  Today, the vast majority of NEOs are discovered by visible light surveys such as the Catalina Sky Survey \citep{Larson.2007a} and PanSTARRS (http://pan-starrs.ifa.hawaii.edu/public/).  Their observations result in a measurement of absolute magnitude H; diameter must be inferred using an assumed albedo.  NEO albedos are known to range widely, from $\sim$1-50\% \citep{Stuart.2004a, Mainzer.2011b}.  Almost always, nothing beyond absolute magnitude H is known to narrow the range of possible albedos, so the uncertainty in albedo ranges from extremely dark to very bright.  In this case, the error in diameter estimated from H alone using an assumed albedo is plus or minus a factor of 3-4 when using the relationship \begin{equation}D = \left[\frac{1329\cdot 10^{-0.2H}}{p_{v}^{1/2}}\right],\end{equation} where $D$ is the effective spherical diameter \citep{Fowler.1992a, Bowell.1989a}.  Therefore, the uncertainty in estimated impact energy can be a factor of roughly 20 if diameter is computed using H alone.  If taxonomic type can be ascertained, it can be used to restrict the range of probable albedos, although the correlation between taxonomic type and albedo is not foolproof \citep[e.g.][]{Stuart.2004a, Mainzer.2011e, Mainzer.2012a, Thomas.2011a}.  However, only a small fraction of NEOs become bright enough to be observed spectroscopically, particularly the dark asteroids that are much more difficult to detect with the visible and near-infrared spectroscopy required for taxonomic classification due to their extreme faintness \citep{Mainzer.2011e}.  For most NEOs, only their absolute magnitude and orbits are known, leading to large uncertainty in diameter and impact energy.

Infrared radiometry allows physical parameters such as diameter and albedo to be determined for large numbers of minor planets.  An asteroid's effective spherical diameter $D$ can be found from its emitted flux at thermal wavelengths \citep[e.g.][]{Lebofsky.1986a, Harris.1998a, Tedesco.2002a}.  Diameters derived from thermal infrared measurements are less sensitive to an object's albedo than those derived purely from reflected sunlight.  If high-quality photometry at multiple infrared wavelengths centered near the peak of an asteroid's thermal emission (typically $\sim$10 $\mu$m for $\sim$300 K NEOs) is available that adequately samples its rotational phase, effective spherical diameter can be determined to within $\pm$10\% \citep{Mainzer.2011c, Mainzer.2011d}.  This error in diameter translates to a much smaller error in impact energy than that derived using only H to estimate size.  Although accurate diameters can be computed using infrared measurements alone, the combination of visible and thermal measurements allows for determination of albedo, which yields clues as to whether an object is stony or carbonaceous \citep[e.g.][]{Tholen.1989a, DeMeo.2009a, Mainzer.2011e, Mainzer.2012a}.  Albedo in turn informs the likely range of densities and hence impact energy, to the extent that it can be tied to taxonomic types and compositional information through linkages to either meteoritic parent bodies \citep[e.g.][]{Consolmagno.1998a, Binzel.1993a, Buratti.2013a} or direct measurements of asteroid density \citep[e.g.][]{Carry.2012a, Merline.2002a}.  

NASA's Wide-field Infrared Survey Explorer mission surveyed the entire sky simultaneously in four infrared wavelengths (3.4, 4.6, 12, and 22 $\mu$m; denoted W1, W2, W3, and W4 respectively) with significant improvements in spatial resolution and sensitivity compared to its predecessors \citep[WISE; ][]{Wright.2010a, Cutri.2012a}.  The spacecraft is in a sun-synchronous polar orbit around the Earth that allows for continuous observations near 90$^{\circ}$ solar elongation.  Over the course of its one-year prime mission, the asteroid-hunting portion of the project known as NEOWISE detected and reported radiometrically-derived diameters and albedos for $>$158,000 asteroids, including $\sim$700 NEOs \citep{Mainzer.2011a}.  More than 160 comets were detected; the infrared data have been used to constrain nucleus sizes, particle size distributions, and gas abundances \citep{Bauer.2013a, Bauer.2012a, Bauer.2011a, Stevenson.2012a}.  NEOWISE detections have been used to set limits on the numbers, orbital elements, sizes, and albedos of asteroid populations throughout the inner solar system \citep[e.g.][]{Mainzer.2011b, Mainzer.2012a, Mainzer.2014a, Masiero.2011a, Grav.2011b, Grav.2012a, DeMeo.2014a}.

The WISE mission surveyed the sky 1.2 times until its dual-stage solid hydrogen cryostat was depleted on September 30, 2010.  The mission was extended an additional four months in order to complete the survey of the inner edge of the main asteroid belt.  The solid hydrogen was required to cool the 12 and 22 $\mu$m channels, but the 3.4 and 4.6 $\mu$m HgCdTe detector arrays continued to operate nominally after the depletion of the cryogen.  The telescope optics and focal planes equilibrated near 73.5 K through passive cooling by continuously pointing near zenith.  

Over the course of the four-month post-cryogenic phase of the prime mission, $\sim$13,500 minor planets were detected in the W1 and W2 channels, including 88 NEOs.  Diameters and albedos derived from the observations at 3.4 and 4.6 $\mu$m were compared with those computed using 3.4, 4.6, 12, and 22 $\mu$m data.  Since the shorter wavelengths span only the Wien side of a typical NEO or Main Belt asteroid's blackbody curve, the effective spherical diameters derived from these measurements alone are less precise, but still accurate to within approximately $\pm$25\% \citep{Masiero.2012a, Mainzer.2012a}.    

WISE survey operations were halted on February 1, 2011 after nine months of fully cryogenic operations and four months of post-cryogenic operations.  On February 17, 2011, the WISE spacecraft was then placed into a hibernation state, and communications with it ceased.  In this mode, the telescope was inertially pointed near the north ecliptic pole, with the solar arrays facing the Sun.  Because they viewed the warm Earth for half of each orbit, the telescope and focal planes warmed to $\sim$200 K.   

In order to continue rapidly surveying and obtaining measurements of minor planet physical properties, the WISE spacecraft was brought out of hibernation on October 3, 2013.  Although the solid hydrogen is now depleted, it is possible to radiatively cool the telescope to low enough temperatures that its heat does not significantly affect sensitivity.  Now known as NEOWISE, the mission is expected to continue until 2017.

\section{NEOWISE Reactivation Mission }

\begin{figure}
\figurenum{1}
\includegraphics[width=6in]{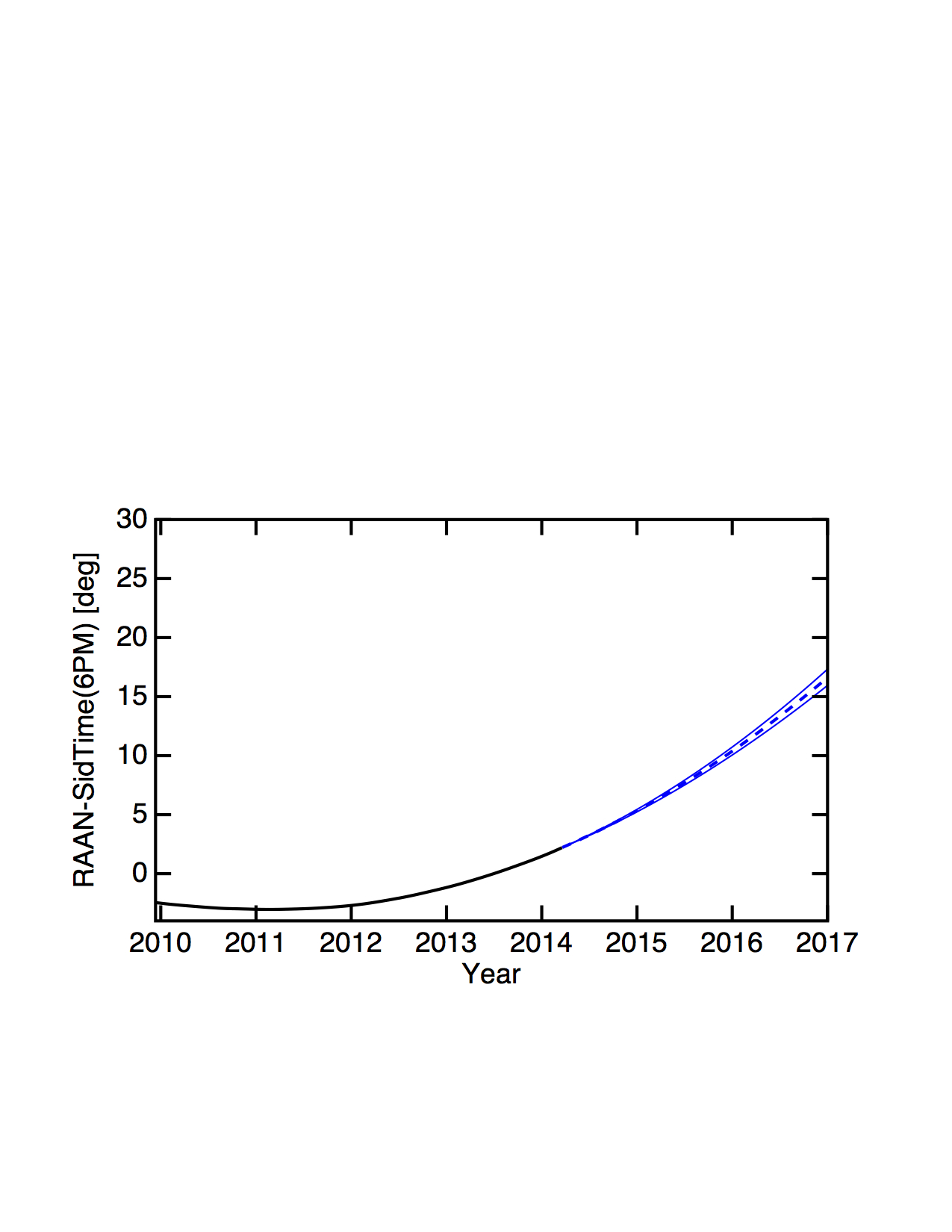}
\caption{\label{fig:DRAAN} Change in right ascension of the ascending node (RAAN) of the WISE spacecraft's orbit as a function of time; measurements are shown as a solid line, and the extrapolated change based on them (bracketed by 2-$\sigma$ errors) is shown as a dashed line.  }
\end{figure}

The scientific objectives of the NEOWISE reactivation mission are to discover and characterize NEOs using its 3.4 and 4.6 $\mu$m channels.  The mission lifetime is limited to $\sim$3 years because the spacecraft's orbital plane is slowly drifting from its ideal Sun-normal orientation under the influence of atmospheric drag (Figure \ref{fig:DRAAN}).  The WISE spacecraft carries no on-board propulsion system, so the rate at which the orbit changes depends solely on the degree to which solar activity affects atmospheric drag forces.  After early 2017, it is anticipated that it will become increasingly difficult to keep light from the Earth and scattered sunlight out of the telescope baffle, bringing a natural end to the mission.  

\begin{figure}
\figurenum{2}
\includegraphics[width=6in]{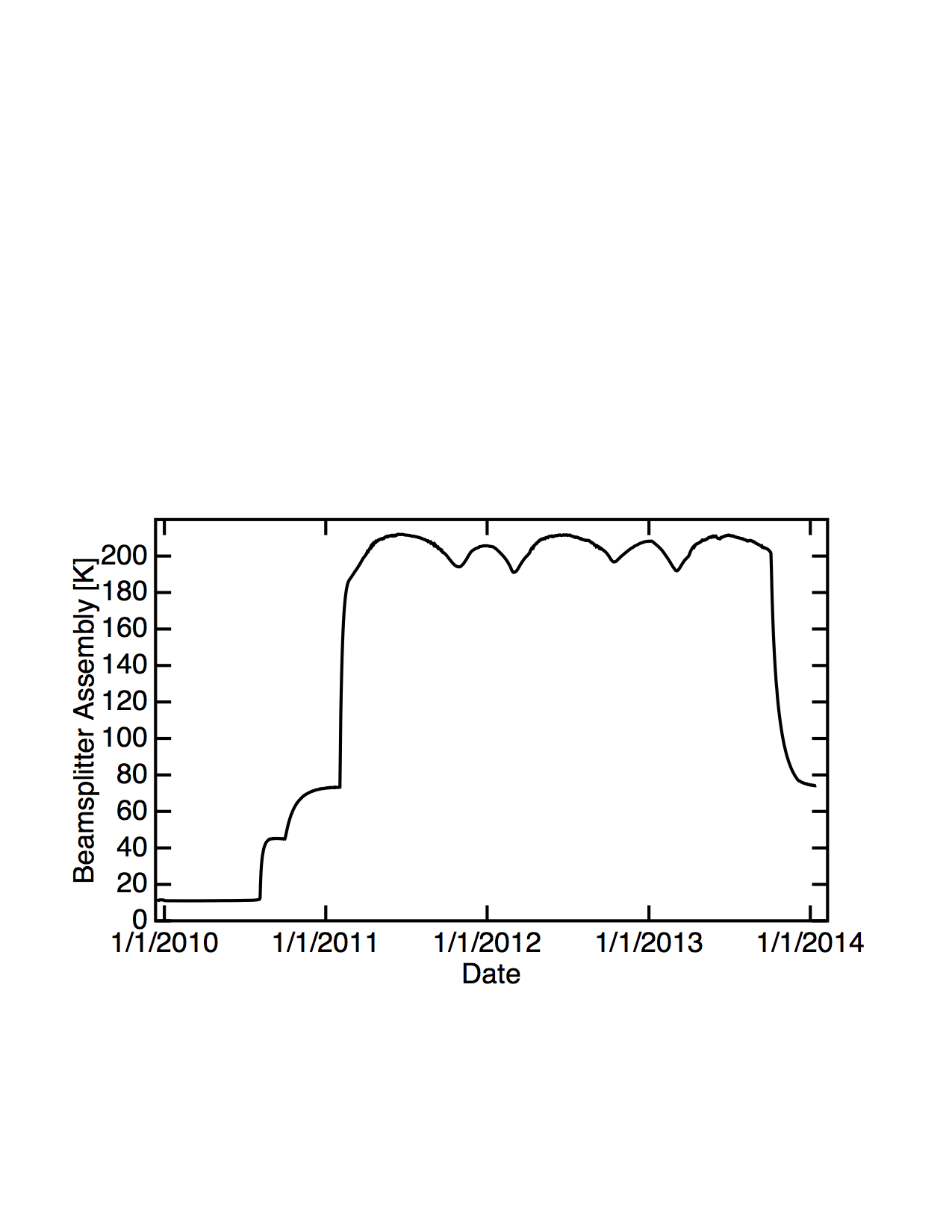}
\caption{\label{fig:cooldown} Temperature of the beamsplitter assembly that holds both W1 and W2 detectors as a function of time.  In the hibernation state, the WISE spacecraft was pointed at the north ecliptic pole; consequently, temperatures throughout the payload, including the telescope structure and focal planes, rose to $\sim$200 K.  Following the start of the NEOWISE survey, the telescope was repointed near zenith, causing the telescope and focal planes to cool via radiation.   }
\end{figure}

\begin{figure}
\figurenum{3}
\includegraphics[width=6in]{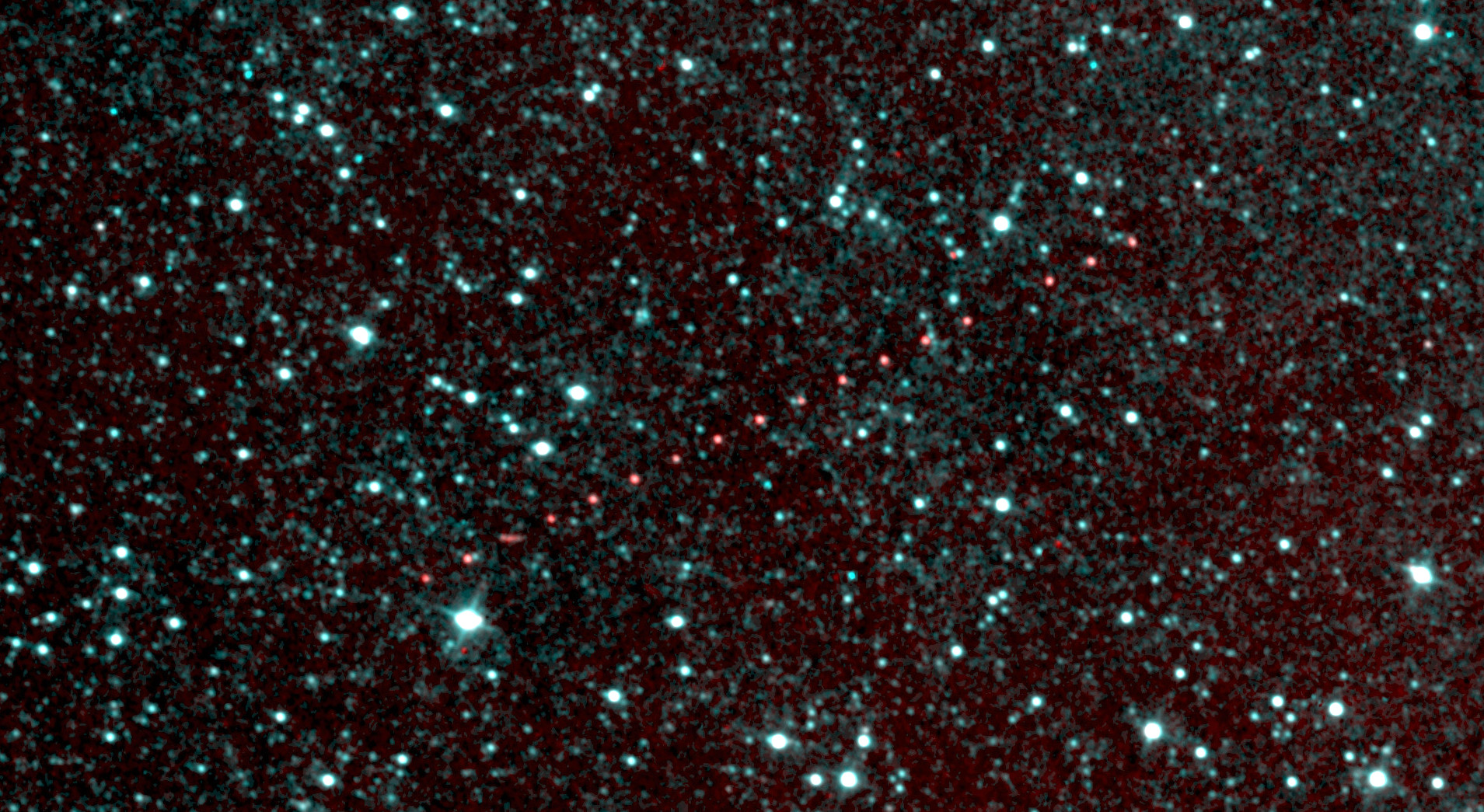}
\caption{\label{fig:holda} A coadded image using some of the first frames collected by the reactivated NEOWISE mission.  Band W1 has been color-coded blue, and W2 is color-coded red.  This coadd was made without outlier rejection to preserve the moving objects.  Main Belt asteroid (872) Holda appears as a string of red dots.  }
\end{figure}

\begin{figure}
\figurenum{4}
\plottwo{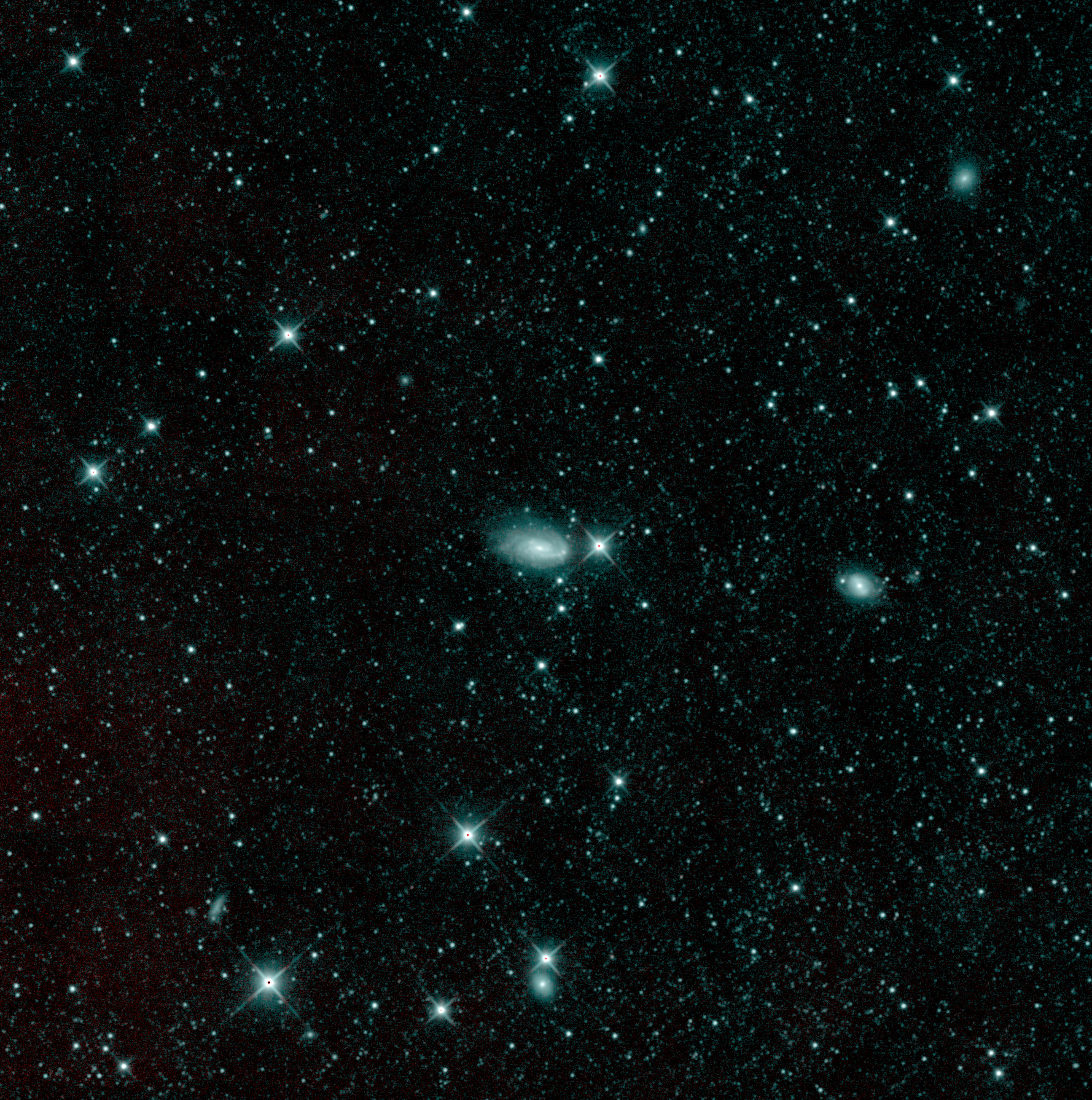}{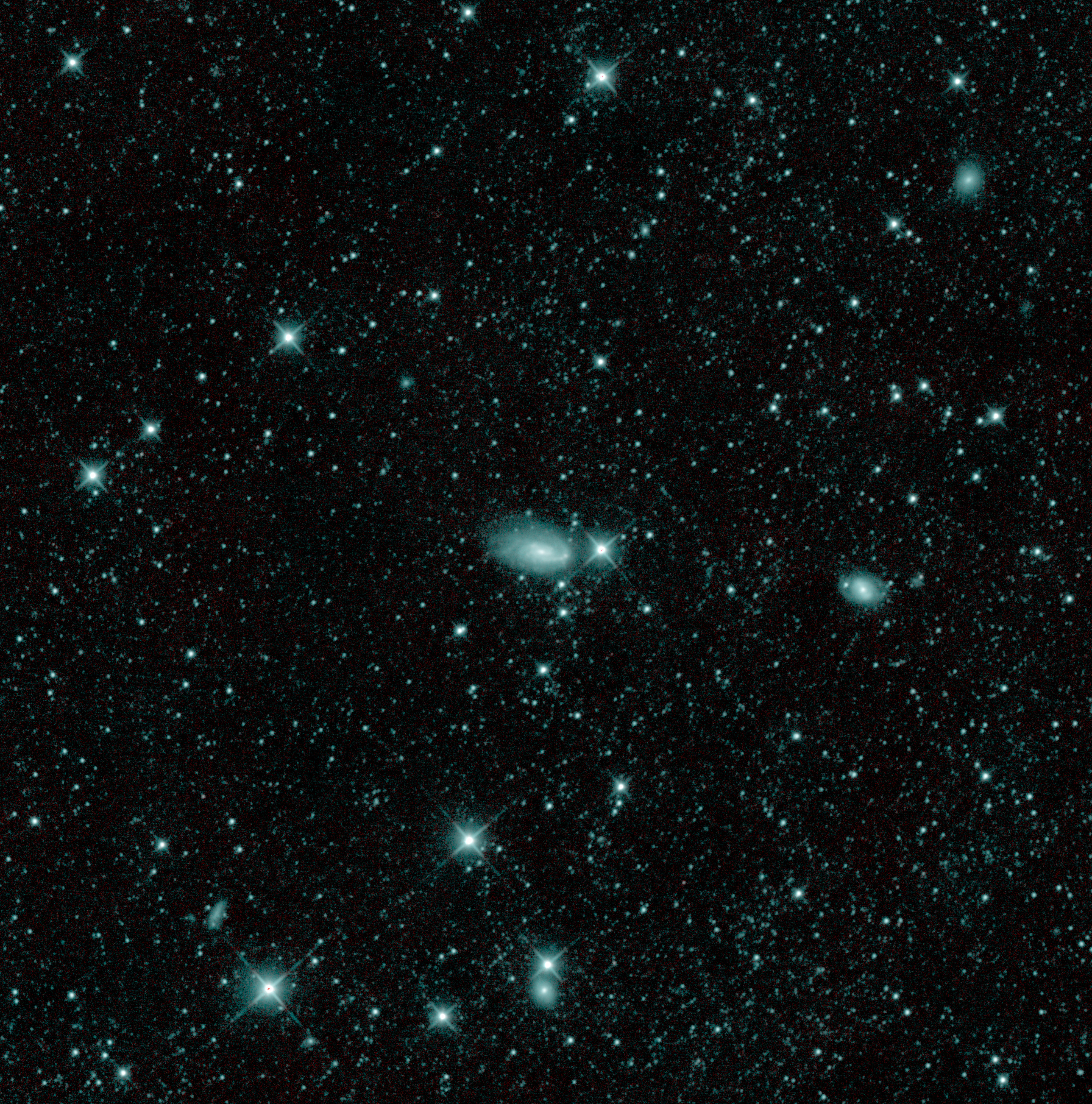}
\caption{\label{fig:first_light} 60 x 60 arc minute images created from two different phases of the mission of the same patch of sky.  Left: Coadded exposures from the prime mission's post-cryogenic phase.  Right: Coadd created from exposures obtained from the reactivated NEOWISE mission. }
\end{figure}

In order to radiatively recool the telescope and detectors, the spacecraft was once again pointed near zenith beginning in October 2013.  After approximately three months, the telescope temperature reached 74 K, completing the cool-down process (Figure \ref{fig:cooldown}).  During this time, the spacecraft's subsystems were checked out, and high-rate communications via Ku-band link to NASA's Tracking and Data Relay Satellite System (TDRSS) and the accompanying ground system in White Sands, New Mexico were reestablished.  The first images from the WISE spacecraft after reactivation were obtained on December 7, 2013 at a telescope temperature of 76.5 K (Figures \ref{fig:holda} and \ref{fig:first_light}).  The flight system began survey operations on December 13, 2013; from December 13 until December 23, a procedure to verify synchronization between the scan mirror and spacecraft scan rate was executed, and science data processing pipeline calibrations were improved.  Regular survey operations, including the moving object processing pipeline, began on December 23, 2013; the first new NEO was discovered six days later.

\begin{figure}[H]
\figurenum{5}
\includegraphics[width=6in]{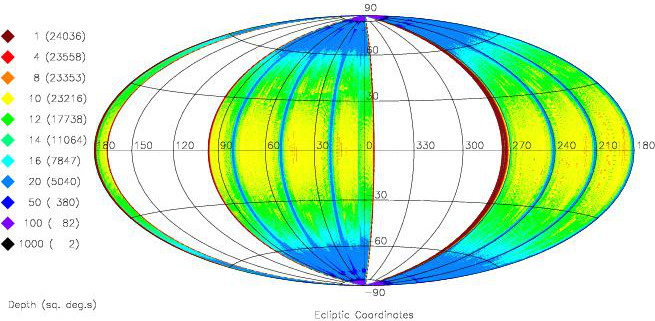}
\caption{\label{fig:coverage} Depth of coverage map achieved by the NEOWISE mission over the first 3 months of survey operations.  The numbers on the left denote the sky area with that depth of coverage or greater, along with the corresponding number of square degrees.}
\end{figure}

The NEOWISE operational cadence remains identical to that employed during the prime mission \citep{Wright.2010a, Heinrichsen.2006a}.  The telescope scans continuously along great circles with approximately constant ecliptic longitude, while a scan mirror freezes the sky on the focal planes for 9.9 seconds and returns to its starting position 1.1 seconds later.  While the sky is fixed on the focal planes, simultaneous exposures are collected in the W1 and W2 bands through the use of beamsplitters every 11 seconds with an exposure time of 7.7 seconds.  The 47x47 arcmin field of view scans at 92.5$^{\circ}$ solar elongation, with 10\% overlaps in the in-scan direction.  The scan path progresses 1 $^{\circ}$/day as the Earth orbits the Sun.  From December 23, 2013 to March 30, 2014, coverage at least one frame deep over 60\% of the entire sky has been achieved (Figure \ref{fig:coverage}).  On average, twelve independent exposures are collected for each point on the ecliptic.  Because the ecliptic poles are densely covered, data downlinks and momentum unloading via magnetic torque rods are always executed near the poles where data loss has minimal impact on science objectives.

As during the prime mission, science data ingest, processing, and archiving is performed at the California Institute of Technology's Infrared Processing and Analysis Center (IPAC) using the data system described in the WISE Explanatory Supplement for the post-cryogenic phase of the mission \citep{Cutri.2012a}.\footnote{http://wise2.ipac.caltech.edu/docs/release/allsky/expsup/sec8\_1.html}  NEOWISE engineering telemetry and science image data are received and merged at IPAC twice per day following downlink from the spacecraft.  The scan/frame pipeline performs basic image calibration, detects and characterizes sources on the individual images, applies astrometric and photometric calibrations, and flags sources that are positionally associated with the expected position of image artifacts.  Identification of moving object candidates does not require the images to be coadded together; hence, only single exposure processing is performed.  The NEOWISE single exposure images and extracted source databases will be publicly released via the IPAC/NASA Infrared Science Archive (IRSA) on an annual basis, starting in March 2015.  

\section{Science Objectives and Preliminary Performance}
Based on its present rate of NEO observations, over the course of its three year mission, NEOWISE is expected to observe $\sim$2000 NEOs, roughly 700-800 of which will be detected in single-exposure images, with the remainder being recoverable through stacking.  Since the observing cadence results in an average of $\sim$12 detections spaced evenly over $\sim$1.5 days, it is possible to recover many objects just below the single-exposure detection threshold by creating comoving stacks of images of previously known objects \citep[c.f. the methods described in ][]{Masci.2009a, Masci.2013a, Mainzer.2014a}.  As with the prime mission, NEOWISE is particularly effective at discovering dark NEOs that are preferentially missed by visible light surveys.  Radiometrically-derived diameters and albedos for all minor planets detected during the survey will be delivered to NASA's Planetary Data System.  

By virtue of the fact that the spacecraft always observes close to 90$^{\circ}$ solar elongation, 25\% of the NEOs detected by NEOWISE are classified as PHAs.  Although ground-based surveys dominate discoveries of PHAs, they constitute a decreasing fraction of their NEO discoveries; in 2013, 8\% of NEOs discovered by ground-based surveys were PHAs.  The NEOs and PHAs observed by NEOWISE during the post-cryogenic phase of the prime mission tended to be large, with a median diameter of $\sim$800 m.  Unlike visible light surveys, which have difficulty detecting low albedo objects, the NEOWISE discoveries tended to be darker than the populations discovered by other surveys \citep{Mainzer.2011b}.

During the prime mission's post-cryogenic phase, the first known Earth Trojan, 2010 TK$_{7}$, was discovered \citep{Connors.2011a}, along with another long-lived Earth co-orbital, 2010 SO$_{16}$ \citep{Christou.2011a}.  As an Earth Trojan librating around the L4 Lagrange point, 2010 TK$_{7}$ spends most of its time in the daylight sky and was  discovered because WISE observes continuously near the twilight-dawn skies where ground-based surveys can spend only a little time.  The asteroid's libration only carries it as far as $\sim$95$^{\circ}$ solar elongation.  The question as to whether or not 2010 TK$_{7}$ represents the first of a larger population of Earth Trojans remains an open one, along with how long ago such a population might have been trapped into resonance with the Earth \citep{Tabachnik.2000a}.  By continually surveying near 92.5$^{\circ}$ solar elongation for three years, the mission will set significantly stricter limits on the population of Earth co-orbitals such as 2010 TK$_{7}$ and 2010 SO$_{16}$.  

Over the course of the three-year NEOWISE mission, the entire sky will be observed six times at 3.4 and 4.6 $\mu$m.  Tens of thousands of Main Belt asteroids and Jovian Trojans will be detected.  The data will also enable a wide range of studies for transient phenomena such as high proper motion, nearby cool stars, variable stars and active galaxies, galactic novae, and supernovae.  

\subsection{Preliminary System Performance}
Instrument performance remains nearly identical to that observed during the prime mission's post-cryogenic phase.  The number of pixels with high dark current has increased slightly, most likely due to radiation effects.  HgCdTe arrays' dark current levels vary strongly with temperature \citep{Beletic.2008a}; since NEOWISE is passively cooled, dark current could change slightly if temperature varies with seasons.  Since the reactivation, 3.3\% and 4.3\% of pixels in bands W1 and W2 have been masked off due to high dark current, compared with 1.9\% and 2.5\% for W1 and W2 during the prime mission's post-cryogenic phase.

\begin{figure}[H]
\figurenum{6}
\includegraphics[width=6in]{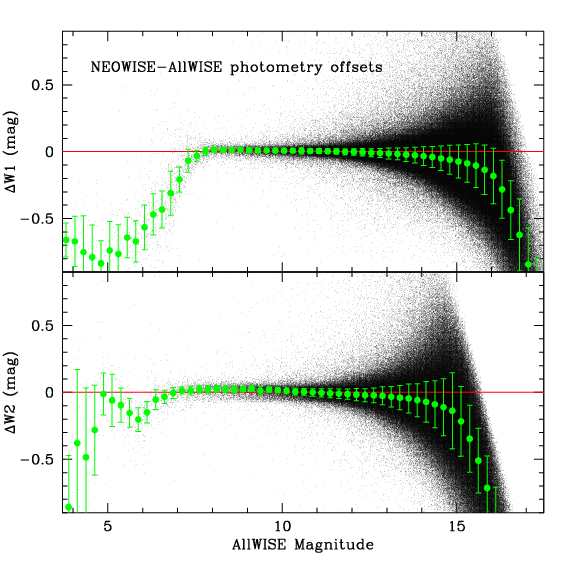}
\caption{\label{fig:photometry} The difference between W1 (top) and W2 (bottom) profile-fit magnitudes measured in all single-exposures in one reactivated NEOWISE scan and those from the AllWISE Source Catalog, plotted as a function of AllWISE Catalog magnitude.  Black dots are individual sources.  Green filled circles and error bars are the trimmed average and RMS of the differences for all sources in 0.25 mag wide bins. }
\end{figure}

Figure \ref{fig:photometry} shows a comparison of the W1 and W2 profile fit photometry from a set of two NEOWISE single exposures compared with the same objects in the much deeper AllWISE Catalog \citep{Cutri.2014a, Cutri.2014b}.  There are essentially no shifts in zero points for either band or flux-dependent biases over most of the brightness range.  The dramatic flux overestimation of NEOWISE fluxes brighter than the saturation levels are consistent with what was observed during the prime mission's post-cryogenic phase.  

The systematic overestimation of NEOWISE fluxes relative to the deeper AllWISE Catalog at the faint end of the distributions is the well-known Eddington bias \citep{Eddington.1913a} that affects measurements of objects near the signal-to-noise detection threshold of any survey.  Faint objects will be preferentially detected when measurement noise drives their apparent brightness above the signal-to-noise threshold.  The same objects will not be detected if negative noise excursions drive their brightness down below the threshold.  This statistical effect can be thought of as an asymmetric truncation of the natural distribution of measured values in the presence of random noise.  Measurements of the truncated distribution will be biased toward positive values, and the amplitude of this bias increases with decreasing signal-to-noise, as shown in Figure \ref{fig:photometry}.  This effect was described in the Explanatory Supplement for the \textit{Infrared Astronomical Satellite} \citep[IRAS;][]{Beichman.1988a}.  If the Eddington bias is not accounted for, fluxes of faint sources will be overestimated.  For example, in a survey in two bands with different detection limits, failure to account for the Eddington bias could result in an erroneous prediction of color changes as a function of asteroid size.  For asteroids detected at thermal infrared wavelengths, the effect is most problematic for faint objects with only a small number of detections and usually results in an overestimate of size.  However, most NEOWISE-detected minor planets have $\sim$10-12 observations spaced evenly over $\sim$36 hours.  An estimate of the systematic effects of the bias in asteroid size estimates made from sparse measurements can be found in \citet{Mainzer.2014a}.

\begin{figure}
\figurenum{7}
\includegraphics[width=6in]{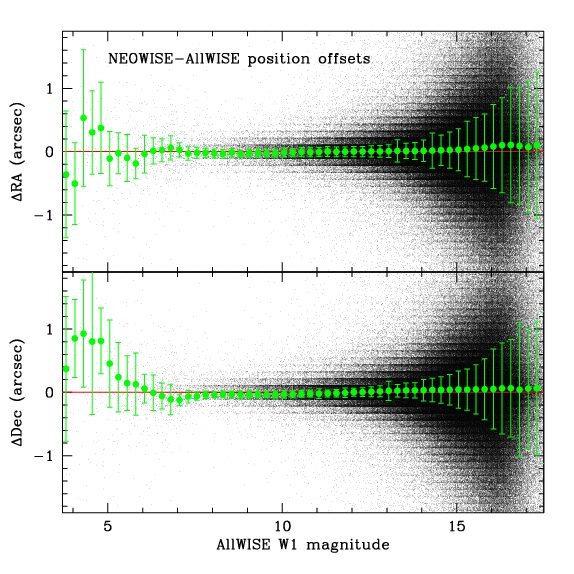}
\caption{\label{fig:astrometry} The differences between reconstructed right ascension (top) and declination (bottom) positions for sources in one reactivated NEOWISE scan.  Color coding is the same as in Figure \ref{fig:photometry}. }
\end{figure}

The very slight systematic increase in $\Delta$W2 in Figure \ref{fig:photometry} over the range $7<W2<14$ mag is also consistent with what was observed during the prime mission's post-cryogenic phase. The deviation is 0.03-0.04 mag, with a mean near zero.  Systematic changes in W1 are $<$0.01 mag.  Figure \ref{fig:astrometry} compares the measured right ascension and declination between NEOWISE and AllWISE sources as a function AllWISE Catalog W1 magnitude; no offsets are observed.

\begin{figure}[H]
\figurenum{8}
\includegraphics[width=6in]{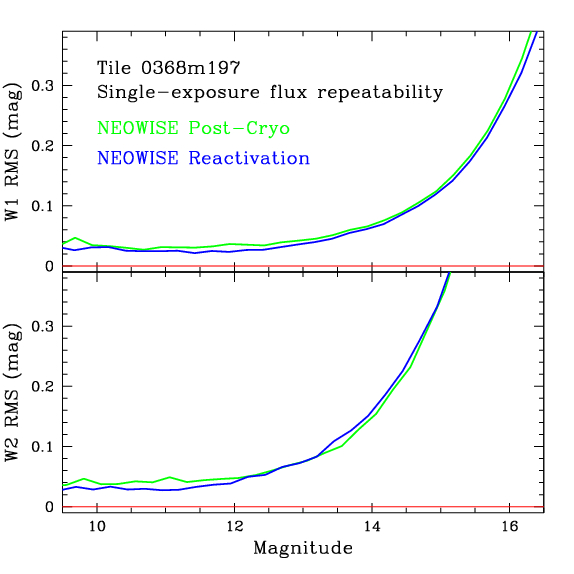}
\caption{\label{fig:repeatability} The RMS of multiple single-exposure flux measurements, in magnitude units, plotted as a function of W1 (top) and W2 (bottom) source magnitude.  Reactivated NEOWISE measurements are shown in blue, and original post-cryogenic phase measurements are shown in green. }
\end{figure}

\begin{figure}
\figurenum{9}
\includegraphics[width=6in]{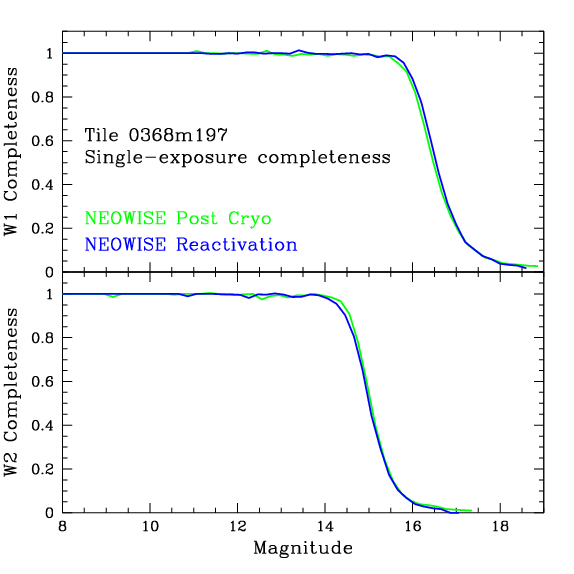}
\caption{\label{fig:nm} The single-exposure detection completeness as a function of W1 (top) and W2 (bottom) for the same objects measured in the reactivated NEOWISE mission (shown in blue), and the original post-cryogenic phase (green).}
\end{figure}

We have compared the sensitivity of exposures obtained from the NEOWISE reactivation with those collected during the prime mission's post-cryogenic phase.  The WISE Multiframe pipeline \citep{Cutri.2012a} was run on a region of sky (Atlas Tile 0368m197) that was fully covered during both the prime mission post-cryogenic phase and the reactivated NEOWISE survey.  The analysis was restricted to sources having $>$10 exposure coverages.  The Multiframe pipeline produces a coadd of all exposures, as well as lists of extracted sources and their associated photometry from the coadd.  Additionally, the Multiframe pipeline also measures the RMS of the fluxes of each source measured on the individual frames, as well as the single-exposure detection statistics.  These give some measure of the photometric measurement accuracy and single-exposure completeness.  Figure \ref{fig:repeatability} shows that the current NEOWISE average photometric repeatability is essentially identical to the prime mission's post-cryogenic phase.  Figure \ref{fig:nm} shows how the source detection completeness on the single-exposure images varies with source brightness.  The completeness is estimated using the repeated observations of the same region of sky obtained using the NEOWISE survey strategy.  The completeness for sources within each 0.2 mag wide brightness bin is computed by forming the ratio of the total number of times all sources within the bin are detected with SNR$>$2 to the total number of times they are observed.  

\subsection{Moving Object Detection}
Moving object candidates are identified in the NEOWISE data in a fashion similar to that performed during the prime mission with a number of improvements incorporated as a result of continued data analysis since the end of survey operations in 2011 \citep{Mainzer.2011a}.  The system is collectively known as the WISE Moving Object Processing System (WMOPS).  Source lists for a given single exposure are assembled and compared to source lists in overlapping single exposures; sources that are co-located on separate frames within a radius of 5 arcsec are considered stationary and are eliminated from further consideration.  Detections of sources that remain after stationary object rejection are removed from consideration if they fall below a flux signal-to-noise ratio (SNR) of 4.5.  Pairs of detections are linked together into ``tuples", then pairs are linked using the methods of \citet{Kubica.2007a}.  The \citet{Kubica.2007a} method uses hierarchical data structures called k-d trees to recursively partition the sources that could potentially be linked into smaller subsets, with the net result being that the search time is proportional to $\rho log \rho$ (where $\rho$ is the source sky-plane density) instead of $\rho^{2}$.  Links can only be made when detections obey velocity limits of 5$^{\circ}$/day, and adjacent detections (separated by only 11 sec) that fall into the overlap regions between single exposures are not used in the construction of the initial tuple pairs of detections.  The resulting sets of position-time pairs are known as ``tracklets".  A minimum of five detections are required to construct a tracklet in order to ensure that the object is real and not merely a chain of cosmic rays or other noise sources.

\begin{figure}[H]
\figurenum{10}
\includegraphics[width=6in]{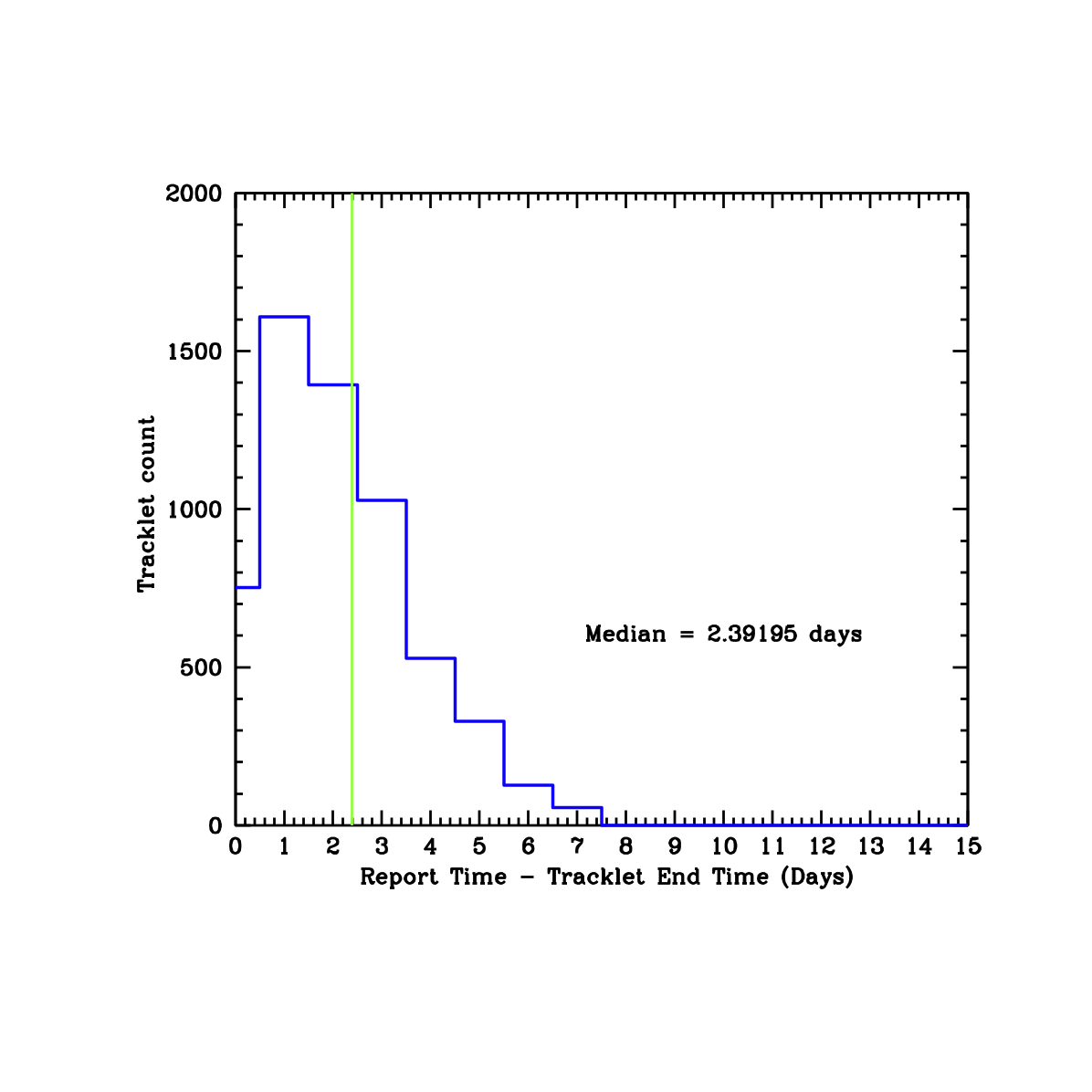}
\caption{\label{fig:lag} The distribution of delivery dates for tracklets from the reactivated NEOWISE mission to the Minor Planet Center, including both previously known minor planets and potential new discoveries.  The mission is required to deliver tracklets within 10 days of observation.}
\end{figure}

Tracklets for which all detections cannot be associated with a previously known solar system object are visually examined by quality assurance astronomers.  An example of an automatically generated quality assurance page is shown in \citet{Mainzer.2011a}.   Tracklets are required to be reported to the International Astronomical Union's Minor Planet Center (MPC) within 10 days of the midpoint of their observation on board the spacecraft.  This requirement ensures that the uncertainty in NEO candidates' ephemerides does not grow beyond $\sim$1$^{\circ}$; uncertainties much larger than this exceed the fields of view of most available follow-up facilities.  With an average observational arc of $\sim$1.5 days, but as little as $\sim$0.4 days, ground-based follow-up is essential to secure NEO candidates' orbits.  As during the prime mission, NEOWISE observations alone are generally of insufficient length to declare NEO candidates officially discovered.  At present, WMOPS is run three times per week.  The average achieved lag time between tracklet endpoint and delivery to the MPC is approximately 2.4 days (Figure \ref{fig:lag}).  

As during the prime mission, follow-up observations of NEOWISE NEO candidates are essential for securing orbits.  Furthermore, visible light observations are necessary for the determination of albedo \citep[e.g.][]{Lebofsky.1989a, Harris.1998a}.  Candidates are placed onto the MPC's public NEO Confirmation Page\footnote{http://minorplanetcenter.net/iau/NEO/toconfirm\_tabular.html}.  Ground-based follow-up is carried out by a network of amateur and professional astronomers around the globe.  Because NEOWISE discoveries are more likely to be dark (Figure \ref{fig:diam_alb_neos}), visible magnitudes can be extremely faint; most of the objects were recovered with V$\sim$21-23 mag.  Given the all-sky observing strategy, targets are often found at high or low declinations, regardless of lunar phase or weather, resulting in unique challenges for follow-up observers.  Follow-up observations are critical to achieving the mission's scientific objectives, and the project greatly appreciates the contributions made by the NEO observing community.  To date, only one NEOWISE NEO candidate out of ten has been lost due to lack of follow-up.  

\section{Preliminary Results}

\begin{figure}[H]
\figurenum{11}
\plottwo{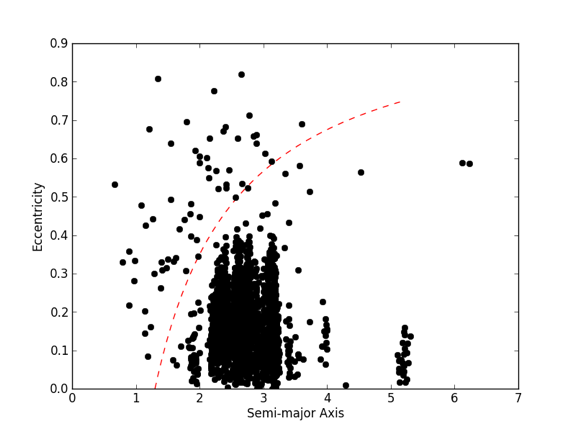}{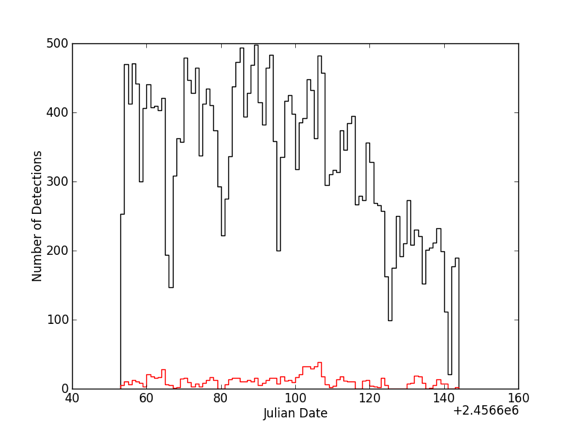}
\caption{\label{fig:WMOPS} Left: Detections as a function of time for all minor planets observed by NEOWISE (black line) and NEOs (red line).  Minor planet detections decrease during the period when the scan circle crosses through the Galactic Center.  Right: The semi-major axis vs. eccentricity of minor planets detected by the reactivated NEOWISE mission from the first month of survey operations; objects to the left of the red dashed line are NEOs.}
\end{figure}

\begin{figure}[H]
\figurenum{12}
\includegraphics[width=6in]{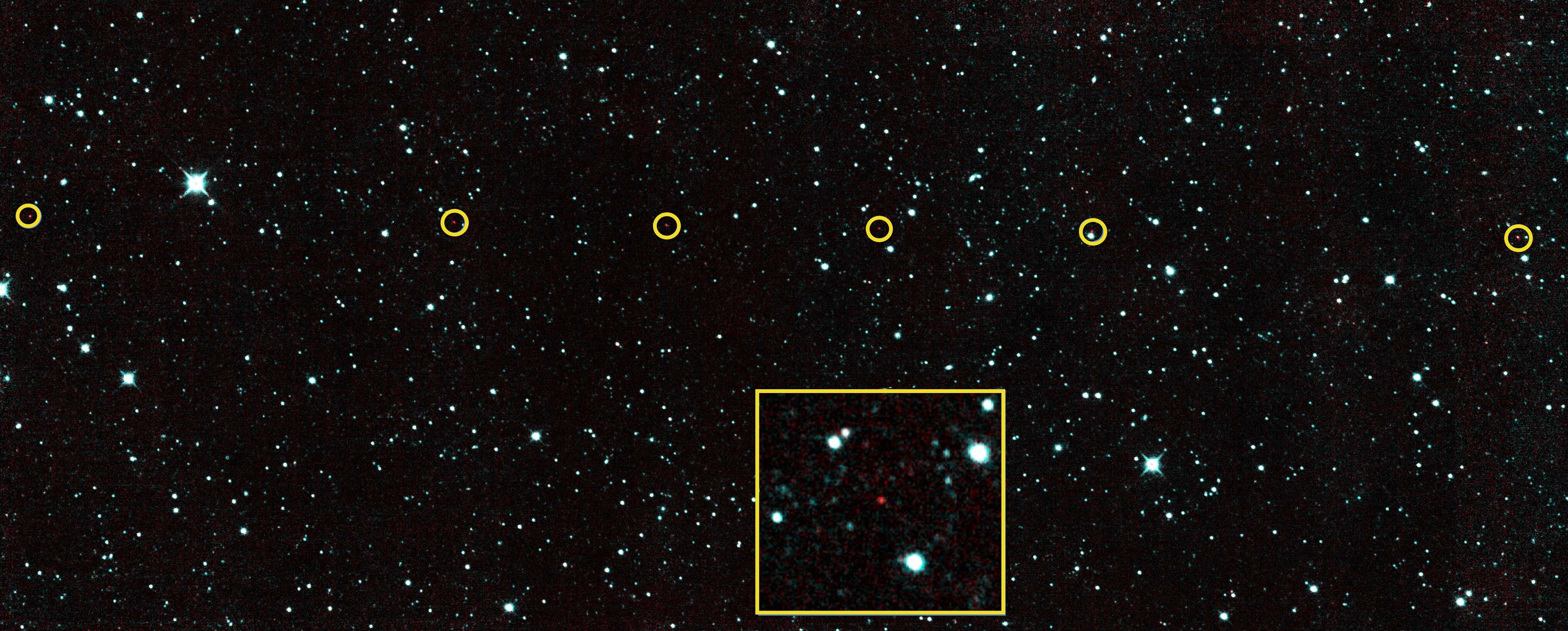}
\caption{\label{fig:2013YP139} The first new NEO discovered by the reactivated NEOWISE mission, 2013 YP$_{139}$ (circled), is revealed to be large, 660$\pm$190 m, and dark.  The inset shows a zoomed-in view of one of the detections.  This image covers $\sim1.5^{\circ}$ of sky; the six detections span 0.4 days.}
\end{figure}

The NEOWISE mission's minor planet detection efficiency is very similar to that achieved during the post-cryogenic phase of the prime mission.  In the 90 days following the survey start on December 23, 2013, WMOPS has recorded detections of 2,915 minor planets, of which 62 are NEOs and ten are comets (Figure \ref{fig:WMOPS}, left).  This number includes ten new NEOs discovered by NEOWISE, along with one new comet, C/2014 C3 NEOWISE, a Halley family retrograde comet discovered on February 14, 2014.  The first new NEO, 2013 YP$_{139}$, was discovered six days after the start of survey operations (Figure \ref{fig:2013YP139}).  All but two of the NEOs discovered by NEOWISE to date have been extremely dark, with albedos lower than $\sim$0.05.  The number of NEO detections, $\sim0.68\pm$0.09/day, is similar to the rate achieved during the post-cryogenic phase of the prime mission (88 unique NEOs detected in four months, or 0.73$\pm$0.11/day).  The NEO detection rate has been lowered due to the effects of confusion in the galactic plane in March 2014 (Figure \ref{fig:WMOPS}, right); NEO detection rates will increase once the scan circle moves past the galactic plane.  Of the thousands of Main Belt asteroids that have been observed to date, $\sim$75\% were also detected during the prime mission; NEOWISE re-detections from different viewing geometries offer the opportunity to perform more detailed lightcurve analyses in order to constrain shapes, rotational states, and thermophysical properties.  

\begin{figure}[H]
\figurenum{13}
\includegraphics[width=6in]{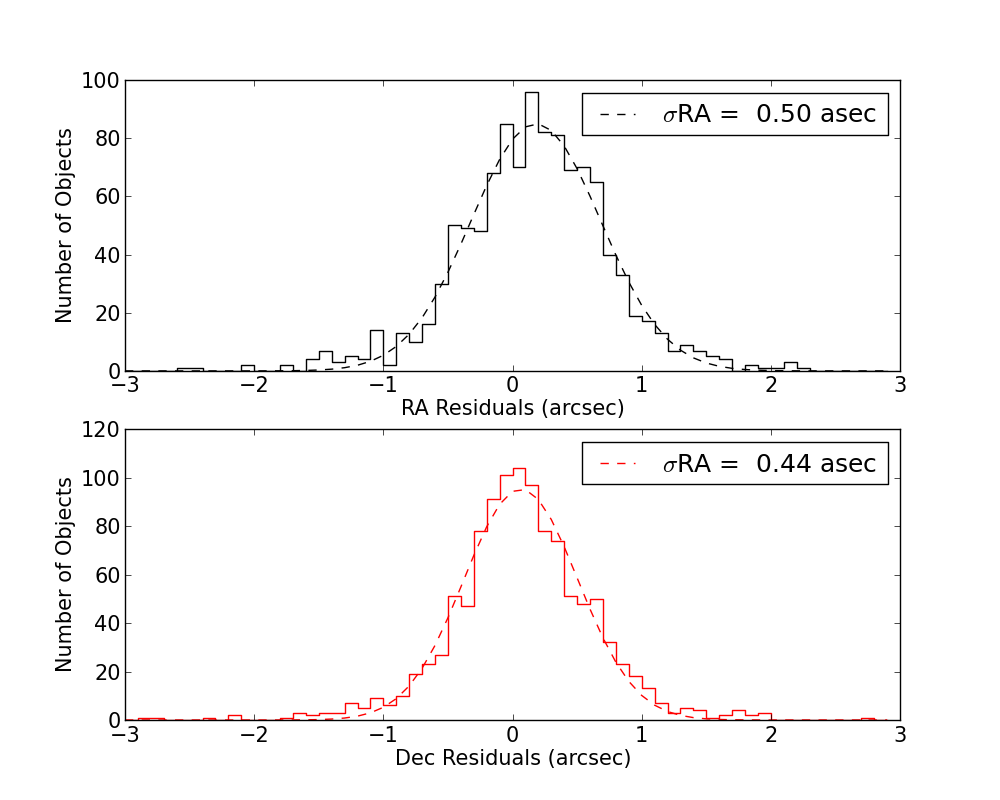}
\caption{\label{fig:residuals} The residual astrometric errors resulting from fits of individual NEOWISE observations to numbered minor planets.}
\end{figure}

It is possible to evaluate the astrometric precision of the NEOWISE observations of minor planets in an independent way by comparing the residual fits to objects with extremely well-known orbits, e.g. numbered objects.  Figure \ref{fig:residuals} shows the astrometric residual errors in right ascension and declination; the root sum square of the medians of these, $\sim$0.67 arcsec, is identical to that observed during all phases of the prime mission.  

\begin{figure}[H]
\figurenum{14}
\includegraphics[width=6in]{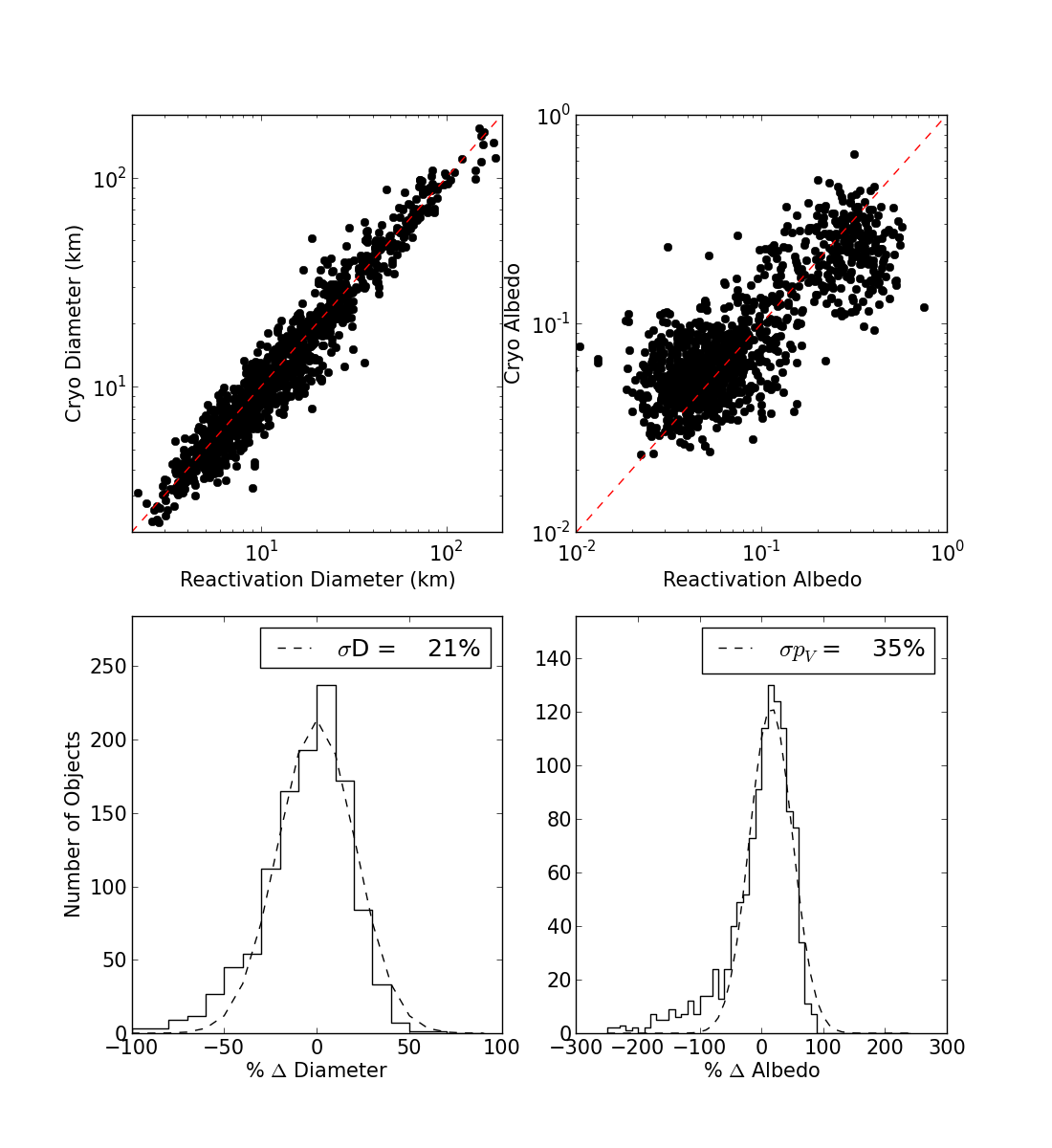}
\caption{\label{fig:diam_alb} Upper left, right: Comparison of effective spherical diameters and albedos for asteroids observed at both 12 and/or 22 $\mu$m during the prime mission and at 3.4 and/or 4.6 $\mu$m during the NEOWISE survey to date.  A one-to-one relationship is shown in both plots as a dashed red line.  Lower left, right: Histogram of differences between 12 $\mu$m and 4.6 $\mu$m NEOWISE reactivation fits for diameter and albedo; Gaussian fits are shown as dashed black lines.}
\end{figure}

Using the Near-Earth Asteroid Thermal Model \citep[NEATM;][]{Harris.1998a} and the methodology described for fitting data from bands W1 and W2 only in \citet{Mainzer.2012c} and \citet{Masiero.2012a}, diameters and albedos computed for objects observed during both the fully cryogenic portion of the prime mission and the NEOWISE reactivation mission can be compared.  The thermal model implementation includes both thermal emission and reflected sunlight; thermal modeling can only be performed if at least one of the bands is thermally dominated.  It was shown in \citet{Mainzer.2011c} and \citet{Mainzer.2011d} that diameters derived from thermal fits performed using WISE 12 and 22 $\mu$m observations produced results accurate to within $\pm$10\% and $\pm$25\%, respectively, when compared to effective spherical diameters and albedos determined from alternate methods such as radar imaging or spacecraft visits. A comparison of thermal fits for $\sim$1200 Main Belt asteroids detected during both the fully cryogenic prime WISE mission at 12 and/or 22 $\mu$m and during the NEOWISE mission at 3.4 and/or 4.6 $\mu$m has been made in order to assess the accuracy of diameters and albedos (Figure \ref{fig:diam_alb}). For these fits, the ratio of the 3.4 $\mu$m albedo to the visible albedo (\irfactor) and the beaming parameter $\eta$ must be assumed; these are taken to be 1.4$\pm$0.5 and 1.0$\pm$0.2 per \citet{Masiero.2012a}.  Uncertainties in H are taken to be $\pm$0.3 mag.  The IAU phase slope parameter G is assumed to be 0.15$\pm$0.10 unless independent measurements were available for a given object from \citet{Warner.2009a} or \citet{Pravec.2012a}; otherwise, H values were taken from the Minor Planet Center.  As with \citet{Mainzer.2011d}, errors are determined through Monte Carlo trials that vary W1, W2, H and G within their respective error bars.  

Figure \ref{fig:diam_alb} shows that diameters agree to within $\pm$21\%, and albedos to within approximately $\pm$35\% of their value (e.g. an object with a 5\% albedo has an uncertainty of $\pm$2\%).  These results are very similar to the post-cryogenic phase of the prime mission, which showed that diameters could be determined to within $\pm$25\% and albedos to within $\pm$40\% \citep{Mainzer.2012c, Masiero.2012a}.  As was observed in these papers, the NEATM tends to underestimate the albedos of dark objects and overestimate the albedos of bright objects.  A possible explanation is dark, carbonaceous asteroids tend to have gray or neutral colors, leading to a lower infrared albedo; \citet{Mainzer.2011e, Mainzer.2012a} found that the mean value of \irfactor\ is closer to $\sim$1.0 for C-types.  Similarly, asteroids with red-sloped visible and near-infrared spectra tend to have higher infrared albedos.  Refitting dark asteroids with \irfactor\ closer to 1.0 increases the resulting albedos slightly.  

It is worth noting that if at least one of the two bands is not thermally dominated, then the data cannot be used for thermal modeling.  All of the NEO thermal fits shown in Tables 1 and 2 are thermally dominated in W2.  \citet{Masiero.2012a} describes the falloff in reflected light versus heliocentric distance and \pIR\ in band W2  for the Main Belt asteroids observed during the post-cryogenic phase of the prime mission.  For dark asteroids, the W2 signal becomes thermally dominated by $\sim$4 AU.  

\begin{figure}[H]
\figurenum{15}
\includegraphics[width=6in]{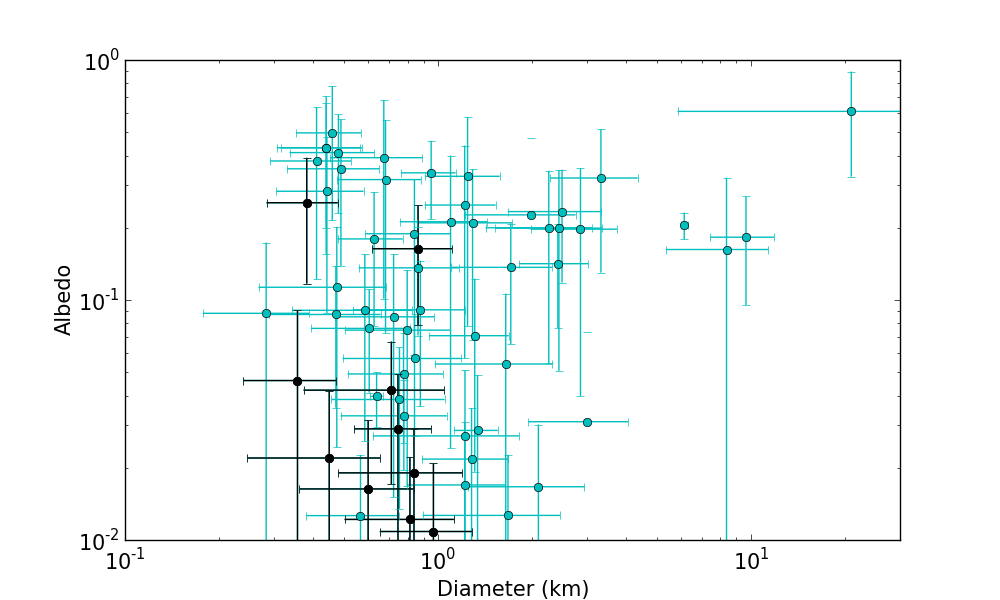}
\caption{\label{fig:diam_alb_neos} Preliminary diameters versus albedos for the NEOs detected by the reactivated NEOWISE mission to date; NEOWISE discoveries are shown as black points.}
\end{figure}

Preliminary thermal fit results for the first 61 NEOs detected by NEOWISE are shown in Figure \ref{fig:diam_alb_neos}.  As was observed during the prime mission, NEOWISE NEO discoveries are usually large (with an average diameter of $\sim$800 m) and low albedo, filling an area of discovery phase space that is harder for ground-based visible light surveys to cover.  These properties of the NEOWISE survey are a result of the wavelength and observing strategy.  The preliminary NEOWISE photometry and thermal fit results for these objects are given in Tables 1 and 2.  

Default values of $\eta=1.4\pm0.5$ and \irfactor=1.6$\pm$1.0 were used following \citet{Mainzer.2012c}, unless $\eta$ and \irfactor\ were determined through other NEOWISE observations using W3 and/or W4 at a similar phase angle from \citet{Mainzer.2011b}.  As described in \citet{Mainzer.2011b, Mainzer.2011d, Mainzer.2011e} and \citet{Masiero.2011a}, the differences in beaming between NEOs and Main Belt asteroids could represent real differences in thermal inertia and temperature distributions between these two populations. However, the phase angles at which NEOs and MBAs are observed by NEOWISE are generally quite different, meaning that more of the night side is observed for NEOs, which can also affect the beaming parameter.  The correspondence between \irfactor\ and taxonomic type for NEOs, MBAs, Hilda-group asteroids, and Jovian Trojans is described in \citet{Mainzer.2011e, Mainzer.2012a, Masiero.2011a, Grav.2012a, Grav.2012b}.

The distribution of bright vs. dark NEOs appears similar to that observed at 12 $\mu$m during the fully cryogenic phase of the mission \citep{Mainzer.2011b}.  Through careful determination of the survey's biases with respect to albedo and orbital elements, it should be possible to extrapolate samples collected by NEOWISE to the larger population \citep[c.f.][]{Mainzer.2011b, Grav.2011a, Grav.2012a}.

\subsection{Comets}
Infrared observations yield information regarding the nucleus, dust and gas of comets.  Nucleus sizes for comets that are unobscured by dust or gas coma can be derived from the thermal component of the infrared flux.  Comets must be closer than $\sim$4 AU in order for the thermal signal from their nuclei to begin to dominate reflected sunlight at 4.6 $\mu$m \citep{Bauer.2013a}, if they are not active.  However, by the time they reach $\sim$3 AU, most comets are likely to become active \citep{Wyckoff.1982a}.  If active, reflected sunlight from the dust and molecular emission can dominate the signal at 4.6 $\mu$m.  For active comets, infrared observations can constrain the distribution of dust, dust temperature, reflectance, production and particle size \citep[c.f.][]{Bauer.2011a, Bauer.2012a, Bauer.2012b}. Infrared observations sample dust that is larger than visible wavelengths and so set more definitive lower bounds on dust mass loss for comets \citep{Bauer.2008a, Bauer.2012a}. 

NEOWISE bands also provide information on gas species produced by comets. A strong CO$_2$ $\nu_{3}$-band emission line at 4.26 $\mu$m falls within the W2 bandpass, so W2 observations enable measurement of CO$_2$ production in comets.  This line is obscured by Earth's atmosphere, but is detectable from space.  Two effects must be mitigated to derive a CO$_{2}$ production rate.  Dust signal must be estimated from the 3.4 $\mu$m band, but W1 provides only an upper bound on the dust signal.  Additional visible wavelength brightness measurements from ground-based telescopes taken near the time of the NEOWISE observations serve to better constrain the dust production.  A weaker CO line at 4.67 $\mu$m falls within the W2 bandpass \citep{Bauer.2011a}; therefore, estimating CO$_{2}$ production requires an assumption that the signal is not dominated by CO.  Such an assumption is not unreasonable, since per molecule, the CO signal is on the order of 11 times weaker.  However, it is not possible to distinguish between the two molecules without further supposition of what the comet's composition is likely to be.  Thus excess W2 signal formally provides a lower bound on the CO$_{2}$ and CO production, while providing an upper bound on one species or another.

\begin{figure}[H]
\figurenum{16}
\includegraphics[width=3in]{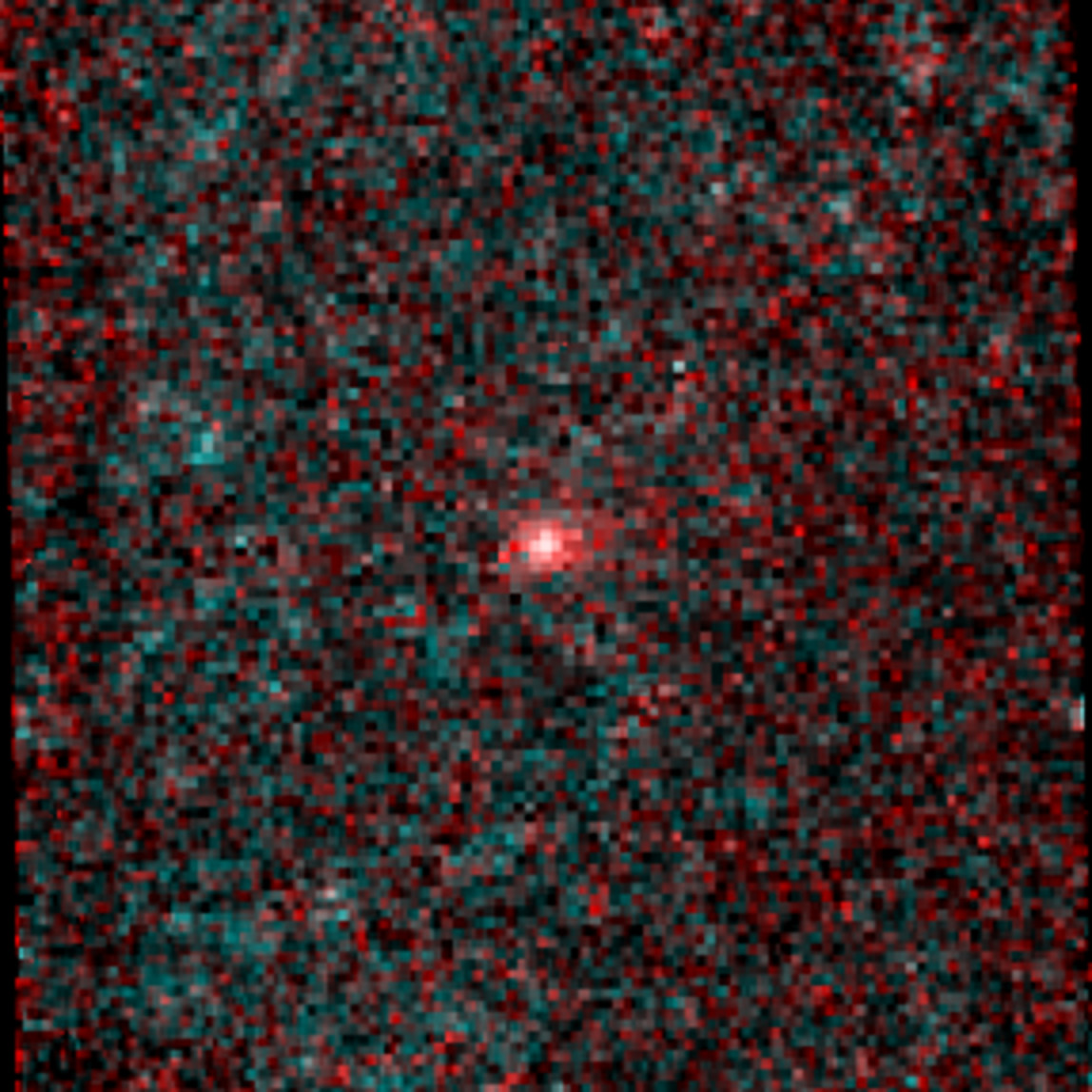}
\caption{\label{fig:CometNEOWISE} A comoving coadd of Comet C/2014 C3 NEOWISE; color-coding is identical to that  in Figure \ref{fig:holda}.}
\end{figure}

To date, ten comets have been detected during the reactivated NEOWISE mission. Half of these objects are long-period comets (LPCs), with orbital periods greater than 200 years. Among them is the comet C/2013 A1 Siding Spring, which is due to pass within 150,000 km of the surface of Mars on October 19, 2014.  Comet C/2014 C3 NEOWISE, a retrograde Halley family comet with a period of 110 years, is the first cometary body discovered during the reactivated NEOWISE mission (Figure \ref{fig:CometNEOWISE}).  At present, cometary activity is initially detected in the NEOWISE data by visual inspection, but automatic detection routines are being evaluated.

\begin{figure}[H]
\figurenum{17}
\includegraphics[width=6in]{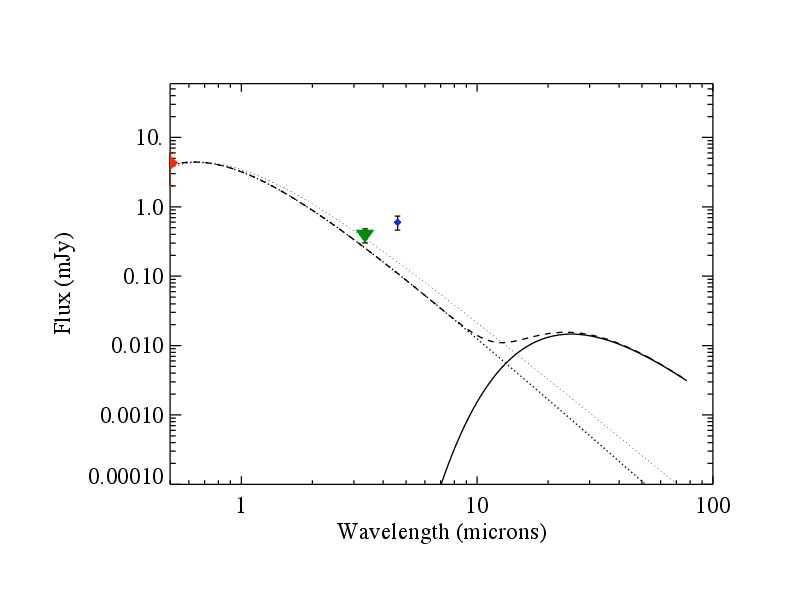}
\caption{\label{fig:comet} The model for the dust in C/2013 A1 Siding Spring is shown; excess W2 emission above the dust signature indicates the presence of CO$_2$ or CO. The dust signal scales as the reflected light flux and thermal flux combined based on $Af\rho$ values in W1 and the visual wavelength measurement (red diamond on the left hand side) reported in \citet{Williams.2014a}. The thermal flux is modeled assuming a blackbody temperature T$_{bb}$ of 146 K appropriate for the comet's heliocentric distance of 3.82 AU at the time of observation. W1 and W2 fluxes are indicated by the green triangle and blue diamond, respectively. The spectrum of reflected light, assuming a neutral spectral response and a particle size distribution (PSD) that scales as the particle size to the -3 power is the heavy dotted line.  The same with reddening law from \citet{Jewitt.1986a} is shown by the dotted line.  A PSD with a -3 power law accounts for the stronger signal at visual wavelengths shown in the plot, since smaller particles do not emit efficiently at longer wavelengths.  The solid line is the thermal flux for T$_{bb}=$146 K for the same projected dust area as indicated by the $Af\rho$ value (log-cm of 2.5). The combined flux for the neutrally reflecting dust is shown by the heavy dashed line.  For each case of dust behavior scaled to the W1 signal, the W2 signal is still greater than can be accounted for by the model dust component. This result yields a W2 excess of $\sim$ 0.3 to 0.5 mJy for an 11 arcsec aperture, and a derived CO$_2$ production of $2.7 \pm 0.5 \times 10^{26}$ mol/sec, assuming all the W2 excess was due to CO$_{2}$, and $2.9 \pm 0.5 \times 10^{27}$ mol/sec if the excess was caused by CO alone.}
\end{figure}

A preliminary analysis of NEOWISE images of C/2013 A1 Siding Spring was conducted on the data obtained from the first month of observation. NEOWISE observed comet Siding Spring in seven exposures in W1 and W2 on January 16, 2014. The SNR measured on each frame exceeded 10 in each band. The W1 and W2 fluxes were $0.4 \pm 0.1 $ and $0.60 \pm 0.14$ mJy, respectively (Figure \ref{fig:comet}).   

$Af\rho$, a measure of dust production in comets, is the product of the dust grain albedo, $A$, the filling factor, $f$, of grains that fall within a circular photometry aperture, and the linear radius, $\rho$, of the aperture at the comet's distance.  $Af\rho$ can be determined from the measured comet fluxes and the comet's known distance \citep{Ahearn.1984a}.  We found $Af\rho$ values of 2.4 to 2.5 log-cm. Assuming grain reflectance values of $\sim 0.04$, similar to those found for other cometary dust grains \citep[c.f.][]{Bauer.2012b}, and grain densities near those of water-ice, the corresponding dust production values were $\sim$ 100 kg/s for ejection velocities on the order of 250 m/s. Assuming a common particle size frequency distribution with log-slope $\sim -3$ with grain radius \citep{Fulle.2004a}, the dust signal within W2 represents only 30-50\% of the total signal.  The derived W2 flux excess can be attributed to CO$_2$ production on the order of $10^{26}$ molecules per second, or a CO production rate of $\sim 10^{27}$ molecules per second. 

\section{Conclusions}
The reactivated NEOWISE mission is conducting a survey with multiple coverages of the entire sky at 3.4 and 4.6 $\mu$m over three years.  Data quality has remained essentially unchanged from the prime mission's post-cryogenic phase, despite 32 months of hibernation.  The data will allow for the characterization and discovery of minor planets and will enable a wide range of time-domain studies.  The mission will result in measurements of radiometric diameters and albedos for $\sim$20\% of the known NEO population over the course of three years.

\section{Acknowledgments}

\acknowledgments{This publication makes use of data products from NEOWISE, which is a project of the Jet Propulsion Laboratory/California Institute of Technology, funded by the National Aeronautics and Space Administration.  This publication makes use of data products from the \emph{Wide-field Infrared Survey Explorer}, which is a joint project of the University of California, Los Angeles, and the Jet Propulsion Laboratory/California Institute of Technology, funded by the National Aeronautics and Space Administration.  We thank the paper's referee for helpful comments that greatly improved the manuscript.  We gratefully acknowledge the services specific to NEOWISE contributed by the International Astronomical Union's Minor Planet Center, operated by the Harvard-Smithsonian Center for Astrophysics, and the Central Bureau for Astronomical Telegrams, operated by Harvard University.  We also thank the worldwide community of dedicated amateur and professional astronomers devoted to minor planet follow-up observations. This research has made use of the NASA/IPAC Infrared Science Archive, which is operated by the California Institute of Technology, under contract with the National Aeronautics and Space Administration. }

\clearpage

\clearpage

\begin{deluxetable}{lllll}
\tabletypesize{\tiny}
\tablecolumns{5}
\tablecaption{\scriptsize{Preliminary NEOWISE magnitudes for the NEOs shown in Figure \ref{fig:diam_alb_neos} at each observation's modified Julian date (MJD).  Objects that were not detected at a particular wavelength represent $2-\sigma$ upper limits \citep{Cutri.2012a}.  A value of ``--" indicates that no data were available at that wavelength.  The final column gives the aperture radius in arcsec used for aperture photometry; ``0" indicates that the pipeline profile fit photometry was used. }  The first ten lines only are shown; the remainder are available in electronic format through the journal website.}
\tablewidth{0pt}
\tablehead{\colhead{Name} & \colhead{MJD} &\colhead{W1 (mag)} & \colhead{W2 (mag)} & \colhead{Aperture}}
\startdata
  01627 & 56676.9179688 & 14.682 $\pm$ 0.080 & 13.280 $\pm$ 0.083 &  0\\
  01627 & 56677.1835938 & 14.236 $\pm$ 0.058 & 13.240 $\pm$ 0.079 &  0\\
  01627 & 56677.3125 & 14.203 $\pm$ 0.114 & 13.190 $\pm$ 0.084 &  0\\
  01627 & 56677.3125 & 14.382 $\pm$ 0.076 & 13.059 $\pm$ 0.074 &  0\\
  01627 & 56677.3789062 & 14.316 $\pm$ 0.060 & 13.370 $\pm$ 0.100 &  0\\
  01627 & 56677.5117188 & 14.379 $\pm$ 0.065 & 13.178 $\pm$ 0.079 &  0\\
  01627 & 56677.578125 & 14.429 $\pm$ 0.069 & 13.249 $\pm$ 0.083 &  0\\
  01627 & 56677.6445312 & 15.603 $\pm$ 0.170 & 14.390 $\pm$ 0.213 &  0\\
  01627 & 56677.7070312 & 14.210 $\pm$ 0.055 & 12.997 $\pm$ 0.073 &  0\\
  01627 & 56677.7734375 & 14.254 $\pm$ 0.059 & 13.376 $\pm$ 0.087 &  0\\
 \enddata
\end{deluxetable}

\clearpage

\begin{deluxetable}{llllllllll}
\tabletypesize{\tiny}
\tablecolumns{10}
\tablecaption{\scriptsize{Thermal fit results for the NEO detections reported in this work. This table contains the preliminary thermal fit results based on the first-pass version of the NEOWISE data processing as described in the text. The columns contain object name, $H$ magnitude, phase curve slope parameter $G$, diameter, visible albedo \pv, beaming parameter $\eta$, infrared albedo $p_{IR}$, and number of observations in each of the two NEOWISE bands.  The $1 -\sigma$ errors presented here were statistically generated using Monte Carlo modeling. NEOWISE magnitudes, absolute magnitude $H$, and $G$ were varied by their $1-\sigma$ error bars, as well as beaming ($\eta$) and $p_{IR}$ when these two parameters could not be fit.    The statistical errors on diameter and \pv\ for each object in the table should be added in quadrature to the  systematic errors described in the text and discussed in \citet{Mainzer.2012a}. }}
\tablewidth{0pt}
\tablehead{\colhead{Name} & \colhead{Name} & \colhead{$H$} &\colhead{$G$} & \colhead{$D$ (km)} & \colhead{\pv}  & \colhead{$\eta$} & \colhead{$p_{IR}$} & \colhead{No. Obs. W1} & \colhead{No. Obs. W2}  }
\startdata
  01627 &       1627 & 12.87 & 0.60 & 8.485 $\pm$ 0.292 &    0.174 $\pm$    0.023 &    1.400 $\pm$ 0.500   &    0.253 $\pm$    0.031 & 13 & 14\\
  02102 &       2102 & 15.90 & 0.15 & 2.494 $\pm$ 0.960 &    0.233 $\pm$    0.177 &    1.400 $\pm$    0.499  &    0.372 $\pm$    0.219 & 16 & 16\\
  03554 &       3554 & 15.87 & 0.15 & 2.419 $\pm$ 0.600 &    0.142 $\pm$    0.065 &    2.120 $\pm$    0.428  &    0.283 $\pm$    0.151 & 22 & 24\\
  04954 &       4954 & 12.75 & 0.15 & 9.494 $\pm$ 0.244 &    0.156 $\pm$    0.023 &    1.400 $\pm$ 0.500   &    0.302 $\pm$    0.061 & 11 & 11\\
  07025 &       7025 & 17.94 & 0.15 & 0.510 $\pm$ 0.171 &    0.452 $\pm$    0.268 &    1.400 $\pm$    0.555  &    0.724 $\pm$    0.273 & 0 & 9\\
  25916 &      25916 & 13.40 & 0.15 & 6.130 $\pm$ 0.159 &    0.205 $\pm$    0.026 &    1.400 $\pm$ 0.500   &    0.234 $\pm$    0.053 & 16 & 20\\
  35107 &      35107 & 16.98 & 0.15 & 1.197 $\pm$ 0.402 &    0.199 $\pm$    0.241 &    1.400 $\pm$    0.475  &    0.319 $\pm$    0.274 & 0 & 16\\
  40267 &      40267 & 16.08 & 0.15 & 1.711 $\pm$ 0.676 &    0.300 $\pm$    0.162 &    1.000 $\pm$    0.164  &    0.481 $\pm$    0.259 & 5 & 5\\
  55532 &      55532 & 16.10 & 0.15 & 1.294 $\pm$ 0.950 &    0.383 $\pm$    0.210 &    1.400 $\pm$    0.863  &    0.448 $\pm$    0.299 & 7 & 7\\
  85182 &      85182 & 17.10 & 0.15 & 1.098 $\pm$ 0.342 &    0.212 $\pm$    0.188 &    1.400 $\pm$    0.423  &    0.339 $\pm$    0.288 & 0 & 9\\
  85628 &      85628 & 16.90 & 0.15 & 0.952 $\pm$ 0.188 &    0.339 $\pm$    0.121 &    2.100 $\pm$    0.451  &    1.000 $\pm$    0.207 & 0 & 11\\
  85774 &      85774 & 18.69 & 0.15 & 1.215 $\pm$ 0.000 &    0.040 $\pm$    0.000 &    1.400 $\pm$ 0.500   &    0.064 $\pm$    0.000 & 12 & 12\\
  89355 &      89355 & 15.74 & 0.15 & 2.009 $\pm$ 0.000 &    0.220 $\pm$    0.000 &    1.417 $\pm$ 0.500   &    0.352 $\pm$    0.000 & 25 & 27\\
  90075 &      90075 & 15.10 & 0.15 & 2.857 $\pm$ 0.879 &    0.197 $\pm$    0.158 &    1.400 $\pm$    0.373  &    0.316 $\pm$    0.323 & 0 & 5\\
  90367 &      90367 & 17.70 & 0.15 & 1.648 $\pm$ 0.671 &    0.054 $\pm$    0.052 &    1.400 $\pm$    0.442  &    0.087 $\pm$    0.130 & 0 & 12\\
  D8127 &     138127 & 17.00 & 0.15 & 0.460 $\pm$ 0.108 &    0.497 $\pm$    0.282 &    1.400 $\pm$    0.375  &    0.497 $\pm$    0.250 & 7 & 7\\
  E2781 &     142781 & 16.10 & 0.15 & 2.262 $\pm$ 0.000 &    0.200 $\pm$    0.000 &    1.400 $\pm$ 0.500   &    0.320 $\pm$    0.000 & 15 & 17\\
  G2080 &     162080 & 19.80 & 0.15 & 0.754 $\pm$ 0.297 &    0.039 $\pm$    0.025 &    1.400 $\pm$    0.466  &    0.062 $\pm$    0.205 & 6 & 6\\
  G3691 &     163691 & 17.00 & 0.15 & 3.000 $\pm$ 1.053 &    0.031 $\pm$    0.042 &    1.400 $\pm$    0.377  &    0.050 $\pm$    0.038 & 0 & 7\\
  I6823 &     186823 & 19.10 & 0.15 & 0.842 $\pm$ 0.346 &    0.057 $\pm$    0.030 &    1.400 $\pm$    0.490  &    0.091 $\pm$    0.220 & 0 & 5\\
  O2450 &     242450 & 14.70 & 0.15 & 3.320 $\pm$ 1.032 &    0.322 $\pm$    0.192 &    1.400 $\pm$    0.457  &    0.516 $\pm$    0.267 & 12 & 12\\
  P0620 &     250620 & 18.10 & 0.15 & 0.864 $\pm$ 0.306 &    0.136 $\pm$    0.065 &    1.400 $\pm$    0.490  &    0.218 $\pm$    0.286 & 0 & 7\\
  Q2623 &     262623 & 18.50 & 0.15 & 0.444 $\pm$ 0.160 &    0.357 $\pm$    0.263 &    1.400 $\pm$    0.564  &    0.571 $\pm$    0.287 & 0 & 6\\
  Q9690 &     269690 & 18.50 & 0.15 & 0.879 $\pm$ 0.342 &    0.091 $\pm$    0.055 &    1.400 $\pm$    0.475  &    0.146 $\pm$    0.221 & 0 & 9\\
  R1480 &     271480 & 17.50 & 0.15 & 0.672 $\pm$ 0.220 &    0.392 $\pm$    0.290 &    1.400 $\pm$    0.507  &    0.627 $\pm$    0.348 & 0 & 7\\
  R6468 &     276468 & 17.90 & 0.15 & 1.312 $\pm$ 0.373 &    0.071 $\pm$    0.052 &    1.400 $\pm$    0.322  &    0.114 $\pm$    0.139 & 0 & 5\\
  U4330 &     304330 & 18.80 & 0.15 & 1.339 $\pm$ 0.214 &    0.029 $\pm$    0.020 &    3.118 $\pm$    0.428  &    0.046 $\pm$    0.032 & 15 & 15\\
  X4673 &     334673 & 17.80 & 0.15 & 0.674 $\pm$ 0.250 &    0.295 $\pm$    0.190 &    1.400 $\pm$    0.505  &    0.472 $\pm$    0.262 & 0 & 13\\
  b7732 &      377732 & 17.00 & 0.15 & 1.291 $\pm$ 0.435 &    0.210 $\pm$    0.142 &    1.400 $\pm$    0.433  &    0.335 $\pm$    0.227 & 6 & 7\\
  c1677 &      381677 & 18.30 & 0.15 & 0.410 $\pm$ 0.118 &    0.380 $\pm$    0.258 &    1.400 $\pm$    0.473  &    0.607 $\pm$    0.220 & 15 & 19\\
  c7733 &      387733 & 19.00 & 0.15 & 0.442 $\pm$ 0.137 &    0.284 $\pm$    0.196 &    1.400 $\pm$    0.491  &    0.987 $\pm$    0.183 & 5 & 6\\
  c9694 &      389694 & 18.20 & 0.15 & 0.440 $\pm$ 0.124 &    0.431 $\pm$    0.231 &    1.400 $\pm$    0.419  &    0.689 $\pm$    0.308 & 5 & 7\\
J98S15B & 1998  SB15 & 21.00 & 0.15 & 0.283 $\pm$ 0.105 &    0.088 $\pm$    0.085 &    1.400 $\pm$    0.477  &    0.141 $\pm$    0.170 & 0 & 15\\
J99S10K & 1999  SK10 & 19.70 & 0.15 & 0.584 $\pm$ 0.242 &    0.091 $\pm$    0.065 &    1.400 $\pm$    0.534  &    0.145 $\pm$    0.195 & 7 & 7\\
K00AK5G & 2000 AG205 & 19.70 & 0.15 & 1.281 $\pm$ 0.392 &    0.022 $\pm$    0.014 &    1.400 $\pm$    0.367  &    0.035 $\pm$    0.056 & 15 & 19\\
K02X40S & 2002  XS40 & 20.10 & 0.15 & 0.779 $\pm$ 0.290 &    0.033 $\pm$    0.013 &    1.400 $\pm$    0.442  &    0.053 $\pm$    0.039 & 14 & 14\\
K03C11C & 2003  CC11 & 19.10 & 0.15 & 1.221 $\pm$ 0.600 &    0.027 $\pm$    0.024 &    1.400 $\pm$    0.512  &    0.043 $\pm$    0.060 & 0 & 17\\
K03G00S & 2003    GS & 19.00 & 0.15 & 0.481 $\pm$ 0.145 &    0.411 $\pm$    0.182 &    1.400 $\pm$    0.487  &    0.661 $\pm$    0.233 & 5 & 6\\
K04M02X & 2004   MX2 & 19.30 & 0.15 & 1.675 $\pm$ 0.781 &    0.013 $\pm$    0.010 &    1.400 $\pm$    0.525  &    0.020 $\pm$    0.045 & 7 & 8\\
K07B00G & 2007    BG & 19.50 & 0.15 & 0.603 $\pm$ 0.209 &    0.076 $\pm$    0.035 &    1.400 $\pm$    0.463  &    0.122 $\pm$    0.228 & 3 & 6\\
K08Q11S & 2008  QS11 & 19.90 & 0.15 & 0.472 $\pm$ 0.195 &    0.087 $\pm$    0.088 &    1.400 $\pm$    0.499  &    0.139 $\pm$    0.141 & 0 & 13\\
K09D01M & 2009   DM1 & 17.10 & 0.15 & 1.711 $\pm$ 0.608 &    0.137 $\pm$    0.071 &    1.400 $\pm$    0.437  &    0.219 $\pm$    0.120 & 49 & 53\\
K09U17X & 2009  UX17 & 21.60 & 0.15 & 0.566 $\pm$ 0.186 &    0.013 $\pm$    0.010 &    1.400 $\pm$    0.416  &    0.020 $\pm$    0.082 & 0 & 15\\
K10L14J & 2010  LJ14 & 17.80 & 0.15 & 0.842 $\pm$ 0.255 &    0.189 $\pm$    0.130 &    1.400 $\pm$    0.407  &    0.302 $\pm$    0.335 & 0 & 22\\
K10O01Q & 2010   OQ1 & 19.00 & 0.15 & 0.722 $\pm$ 0.252 &    0.085 $\pm$    0.070 &    1.400 $\pm$    0.459  &    0.136 $\pm$    0.254 & 0 & 11\\
K11Q48D & 2011  QD48 & 18.20 & 0.15 & 0.440 $\pm$ 0.134 &    0.431 $\pm$    0.275 &    1.400 $\pm$    0.516  &    0.689 $\pm$    0.245 & 5 & 7\\
K13P06X & 2013   PX6 & 18.50 & 0.15 & 2.095 $\pm$ 0.846 &    0.017 $\pm$    0.014 &    1.400 $\pm$    0.429  &    0.027 $\pm$    0.051 & 9 & 10\\
K13W44T & 2013  WT44 & 19.60 & 0.15 & 1.219 $\pm$ 0.424 &    0.017 $\pm$    0.014 &    1.400 $\pm$    0.397  &    0.027 $\pm$    0.036 & 5 & 6\\
K13Y13Z & 2013  YZ13 & 19.60 & 0.15 & 0.475 $\pm$ 0.202 &    0.113 $\pm$    0.141 &    1.400 $\pm$    0.529  &    0.181 $\pm$    0.213 & 0 & 6\\
K13YD9P & 2013 YP139 & 20.60 & 0.15 & 0.968 $\pm$ 0.316 &    0.011 $\pm$    0.010 &    1.400 $\pm$    0.393  &    0.017 $\pm$    0.020 & 6 & 6\\
K14A33A & 2014  AA33 & 19.10 & 0.15 & 0.777 $\pm$ 0.262 &    0.049 $\pm$    0.024 &    1.400 $\pm$    0.385  &    0.078 $\pm$    0.053 & 5 & 5\\
K14A46Q & 2014  AQ46 & 20.30 & 0.15 & 0.839 $\pm$ 0.359 &    0.019 $\pm$    0.010 &    1.400 $\pm$    0.495  &    0.030 $\pm$    0.069 & 0 & 11\\
K14A53A & 2014  AA53 & 19.80 & 0.15 & 0.710 $\pm$ 0.338 &    0.042 $\pm$    0.025 &    1.400 $\pm$    0.566  &    0.067 $\pm$    0.219 & 0 & 14\\
K14B60G & 2014  BG60 & 20.10 & 0.15 & 0.746 $\pm$ 0.206 &    0.029 $\pm$    0.020 &    1.400 $\pm$    0.346  &    0.046 $\pm$    0.145 & 0 & 187\\
K14B63E & 2014  BE63 & 21.20 & 0.15 & 0.599 $\pm$ 0.238 &    0.016 $\pm$    0.015 &    1.400 $\pm$    0.469  &    0.026 $\pm$    0.053 & 0 & 6\\
K14C04Y & 2014   CY4 & 21.50 & 0.15 & 0.450 $\pm$ 0.204 &    0.022 $\pm$    0.020 &    1.400 $\pm$    0.547  &    0.035 $\pm$    0.074 & 0 & 7\\
K14C14F & 2014  CF14 & 17.90 & 0.15 & 0.864 $\pm$ 0.246 &    0.164 $\pm$    0.085 &    1.400 $\pm$    0.398  &    0.262 $\pm$    0.306 & 0 & 6\\
K14D10C & 2014  DC10 & 20.10 & 0.15 & 0.637 $\pm$ 0.028 &    0.040 $\pm$    0.010 &    0.991 $\pm$    0.041  &    0.063 $\pm$    0.016 & 8 & 10\\
K14E00D & 2014    ED & 19.20 & 0.15 & 0.381 $\pm$ 0.098 &    0.254 $\pm$    0.137 &    1.400 $\pm$    0.372  &    0.407 $\pm$    0.266 & 0 & 6\\
K14E45N & 2014  EN45 & 20.85 & 0.15 & 0.814 $\pm$ 0.311 &    0.012 $\pm$    0.010 &    1.400 $\pm$    0.460  &    0.019 $\pm$    0.044 & 0 & 14\\
K14E49Q & 2014  EQ49 & 21.20 & 0.15 & 0.356 $\pm$ 0.117 &    0.046 $\pm$    0.045 &    1.400 $\pm$    0.392  &    0.074 $\pm$    0.079 & 0 & 6\\
\enddata
\end{deluxetable}

\clearpage

\end{document}